\documentclass[aps,prl,letterpaper,10pt,twocolumn,showpacs,superscriptaddress]{revtex4-1}

\usepackage{amsfonts,amssymb,amsmath,amsthm,graphicx}
\usepackage{url}
\usepackage{dsfont}
\usepackage{bbm}
\usepackage[usenames,dvipsnames]{xcolor}
\usepackage{nicefrac}
\usepackage{natbib}
\usepackage{setspace}
\usepackage{subfigure}
\usepackage{comment}
\usepackage[normalem]{ulem} 
\usepackage{todonotes}

\def\bra #1{\langle #1\vert}
\def\ket #1{\vert #1\rangle}
\def\braket #1#2{\langle #1 \vert #2\rangle}

\def\virgolette #1{``#1"}

\begin{document}

\title{Source-device-independent Ultra-fast Quantum Random Number Generation}

\author{Davide G. Marangon}
\affiliation{Dipartimento di Ingegneria dell'Informazione, Universit\`a degli Studi di Padova, Padova, Italia}

\author{Giuseppe Vallone}
\affiliation{Dipartimento di Ingegneria dell'Informazione, Universit\`a degli Studi di Padova, Padova, Italia}
\affiliation{Istituto di Fotonica e Nanotecnologie, CNR, Padova, Italia}

\author{Paolo Villoresi}
\affiliation{Dipartimento di Ingegneria dell'Informazione, Universit\`a degli Studi di Padova, Padova, Italia}
\affiliation{Istituto di Fotonica e Nanotecnologie, CNR, Padova, Italia}



\begin{abstract}
 Secure random numbers are a fundamental element of many applications in science, statistics, cryptography and more in general in security protocols. We present a method that enables the generation of high-speed unpredictable random numbers from the quadratures of an electromagnetic field without any assumption on the input state. The method allows to eliminate the numbers that can be predict due the presence of classical and quantum side information. In particular, we introduce a procedure to estimate a bound on the conditional min-entropy based on the Entropic Uncertainty Principle for position and momentum observables of infinite dimensional quantum systems. By the above method, we experimentally demonstrated the generation of secure true random bits at a rate greater than 1 Gbit/s.
\end{abstract}
%
\maketitle

\emph{ Introduction -}
Quantum Random Number Generators (QRNG)  exploits intrinsic probabilistic 
quantum processes to generate true random numbers. Indeed, 
the expression \virgolette{QRNG} was first introduced for a device 
based on the decay of radioactive nuclei \cite{Schmidt1970}.
Afterwards, QRNGs exploiting the versatility of light were realized:
such devices are based on optical processes such as photon \emph{welcher weg}
\cite{Rarity1994,Jennewein2000,stefanov2000optical}, photon time of arrival
\cite{Furst2010,stipvcevic2007quantum,Wayne2010} or vacuum 
quadratures \cite{Trifonov2007,Gabriel2010, Shen2010, Symul2011}.

Usually, the assessment of the randomness of the generated numbers is obtained
by applying statistical tests on the output bits. In most of the QRNGs, 
passing the tests is the only method used to certify the randomness.
In case of failure (attributed to hardware problem since the process is 
assumed to be \virgolette{random}), numbers are algorithmically 
post-processed until the tests are passed.

However, this procedure con only certifies that the numbers are identically and independently distributed (i.i.d.) with respect to \emph{those} \footnote{In fact, universal tests of randomness do not exist.} applied tests.
Indeed, \emph{a posteriori}
statistical tests 
cannot certify that the numbers are not known to someone possessing side information about the generator.
For instance, it is not possible to eliminate hardware noise, which is a source of classical side information for an eavesdropper, Eve, who may be able to control it. Hence, a statistical test \emph{a posteriori} cannot establish whether the numbers are originated by the quantum process or by the noisy hardware. Moreover, even assuming a QRNG with an ideal noiseless hardware, a statistical test cannot reveal whether the outputs arise from a a quantum measurements and
then are intrinsically random. 
For instance, a polarization welcher weg QRNG with an optical source emitting photons in a completely mixed polarization state can be seen as the photonic version of a fair coin.
The random sequence can be predicted by Eve if she knows ``the coin's equations of motion'' (namely she has classical side information) or if she holds a quantum system correlated with the QRNG (namely she has quantum side information).

The quantity that evaluates the amount of 
side information on a random
sequence $Z$ 
is the so called 
 conditional quantum min-entropy $H_{\text{min}}(Z|E)$ \cite{Robert}. However, such min-entropy is generally hard to estimate.
For instance, in the Device Independent (DI)
framework, $H_{\text{min}}(Z|E)$ can be related to violation of a Bell's inequality. However, these protocols are very 
demanding from the experimental point of view
since they require a loohole-free Bell tests~\cite{Pironio2010b,Colbeck2012a,Gallego2013}.

In this work we propose and experimentally realize an efficient protocol for the secure and ultra-fast generation of random numbers
{able to evaluate a lower bound for the conditional quantum min-entropy.}
The method assumes a trusted measurement device and a complete untrusted source, i.e. a {\it source-device-independent} (SDI) scenario. 
 
The bound is estimated by exploiting the \emph{entropic uncertainty principle} for infinite dimensional quantum systems, by adopting the method introduced in \cite{vallone2014quantum} for the finite dimension case.
We note that, similarly to the protocol introduced in  \cite{vallone2014quantum}, the present scheme does not require any assumption on the dimension of the Hilbert space of the source.
Fast generation rate is provided by a continuous variable (CV) scheme based on the  measurement of quadrature observables of the electromagnetic field.

\begin{figure*}[t]
\includegraphics[width=0.9\textwidth]{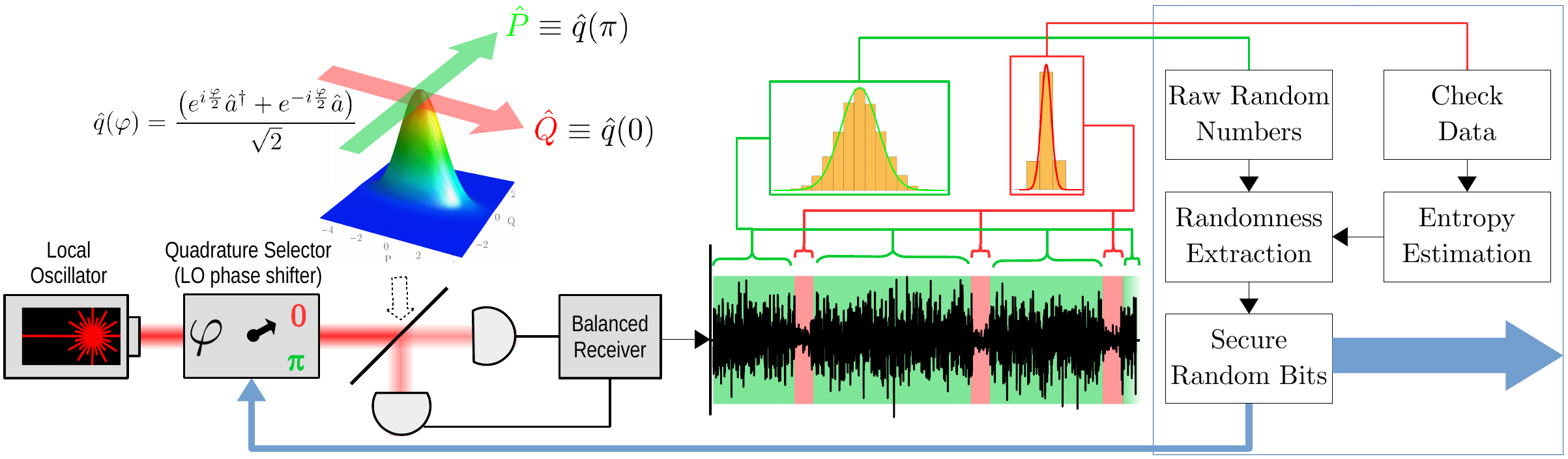}
\caption{Random numbers are obtained by measuring the position quadrature of a Gaussian state with optical homodyne. These numbers are \virgolette{secured} by applying a strong randomness extractor calibrated on a conservative bound of $ H_{\text{min}}(P_{\delta p}|E)$. Such bound is obtained by randomly measuring the complementary quadrature, i.e. the momentum one. Part of the secure bits are \virgolette{re-invested} in the process to sustain the random quadrature switching.}
\label{setup}
\end{figure*}

\emph{Review of CV-QRNG -} In a typical scenario, a CV-QRNG user (Alice) generates random numbers by measuring the momentum quadrature $\hat P$ of a quantum state  $\rho_A$ (typically the vacuum) of an electromagnetic field mode.
With CV systems, the finite resolution of the experimental devices leads to a \virgolette{discretization} of the measurements
(see Supplementary Informations (SI) for more details). More specifically, a coarse grained version of a quadrature operator,  e.g. $\hat P_{\delta p}$, can be obtained by introducing a partition $\mathcal P_{\delta p}= \{I_{\delta p}^k \}$ of its possible output values $p\in\mathbb{R}$. In the above expression, $I_{\delta p}^k$ are given by the half-open intervals 
$I_{\delta p}^k =\left(k\delta p,(k+1)\delta p\right]$ with $k\in\mathbb N$ and $\delta p$ the \emph{precision} of the measurement. Alice measures the POVMs $\{\hat{P}^k_{\delta p}\}$ with elements $\hat{P}^k_{\delta p}=\int_{I^k_{\delta p}}dp\ket{p}\bra{p}$ and stores the outcomes $p_k$ appearing with probability $\mathfrak{p}(p_k)={\rm Tr}[\rho_A \hat{P}^k_{\delta p}]$ in a classical register $P_{\delta p}$.

For cryptographic applications Alice needs to evaluate the probability $\mathfrak{p}_{\text{guess}}(P_{\delta p}|E)=2^{-H_{\text{min}}(P_{\delta p}|E)}$, that an adversary (Eve) has to guess correctly the outcome of a measurement by adopting an optimal strategy. The guessing probability depends on the \emph{quantum conditional min-entropy} $H_{\rm min}(P_{\delta p}|E)$, which represents the maximal content of true random bits achievable for each measurement from the system  $A$, i.e. uniform and uncorrelated from any classical or quantum side-information held by an eavesdropper \cite{Colbeck2012a,Frauchiger2013}.

Previous works on CV-QRNGs assumed that the state $\rho_A$ is pure \cite{Gabriel2010}. In this case, the conditional min-entropy reduces to the  \emph{classical min-entropy} $H_{\infty}(P_{\delta p})=-\log_2\left[\max_k \mathfrak{p}(p_k)\right]$. Eve's best strategy consists in betting on the most probable value, namely $\mathfrak{p}_{\rm guess}(P_{\delta p})=\max_k{\rm Tr}\left[\rho_A\hat{P}^k_{\delta p}\right]$, according to the Born rule. Other works assumed the eavesdropper intrusion limited to the  {classical  noise \cite{Gabriel2010,Symul2011}
which unavoidably affects the experimental apparatus.} 
However, 
to generate true randomness it is necessary to  consider also quantum side information: indeed,
the most general scenario contemplates the possibility of an eavesdropper 
having access to a quantum system $E$ correlated with the system $A$. It is worth to stress that such scenario is not \emph{paranoid} but it results from relaxing the \emph{strong} assumption of the system $A$ 
being in a pure state. 

\emph{SDI-CV random number generator -}
In the untrusted source scenario, the state $\rho_A$ is in general \emph{mixed}: it can be purified by a state $\rho_{AE}$, namely $\rho_A=\text{Tr}_E\left[\rho_{AE}\right]$ where $E$ can be identified with the already mentioned eavesdropper, or with the system ``environment".
We note that the mixedness of $\rho_A$ corresponds to  common physical situations: any decoherence or imperfection in the state preparation leads to correlations with the environment $E$.
In this general case, Alice can estimate the exact value of $H_{\text{min}}(P_{\delta p}|E)$ only by performing a complete quantum state tomography. 

However, an alternative and simpler 
approach consists in estimating a lower bound. This can be obtained by exploiting  the \emph{entropic uncertainty principle} (EUP) for conditional min- and max-entropies in the presence of infinite dimensional quantum memories introduced by Furrer \emph{ et al.} \cite{Furrer2014}. The EUP can be summarized as follows: let's consider a tripartite state $\omega_{ABE}$  with  Alice, Bob  and  Eve holding infinite dimensional quantum systems $A$, $B$ and $E$ respectively. Alice measures quadratures $\hat P_{\delta p}$ and $\hat Q_{\delta q}$ on  $\omega_A=\text{Tr}_{BE}\left[{\omega_{ABE}}\right]$ and she stores the outcomes in two classical systems $P_{\delta p}$ and $Q_{\delta q}$. The EUP is written as
\begin{equation}
\label{EUP}
 H_{\text{min}}(P_{\delta p}|E)+H_{\text{max}}(Q_{\delta q}|B)\geq - \log_2 c(\delta q,\delta p)\,,
\end{equation}
where
\begin{equation}\label{eq:QPconstDisc}
c(\delta q,\delta p) =\frac{1}{2\pi}\delta q\delta p\cdot S_{0}^{(1)}\left(1,\frac{\delta q\delta p}{4}\right)^{2}
\end{equation}
and $S_{0}^{(1)}$ is the $0^{\text{th}}$ radial prolate spheroidal wavefunction of the first kind \cite{landau1961prolate}. In Eq. \eqref{EUP}, $H_{\text{min}}(P_{\delta p}|E)$ quantifies Eve's uncertainty about the outcomes $p_k$ , while the conditional max-entropy $H_{\text{max}}(Q_{\delta q}|B)$ expresses Bob's lack of knowledge about $q_k$. The term $c(\delta q,\delta p)$ is the \virgolette{incompatibility} of the measurement operators, i.e. it is maximal if the operators are maximally complementary. 

\begin{figure}[t!]
\includegraphics[width=0.22\textwidth]{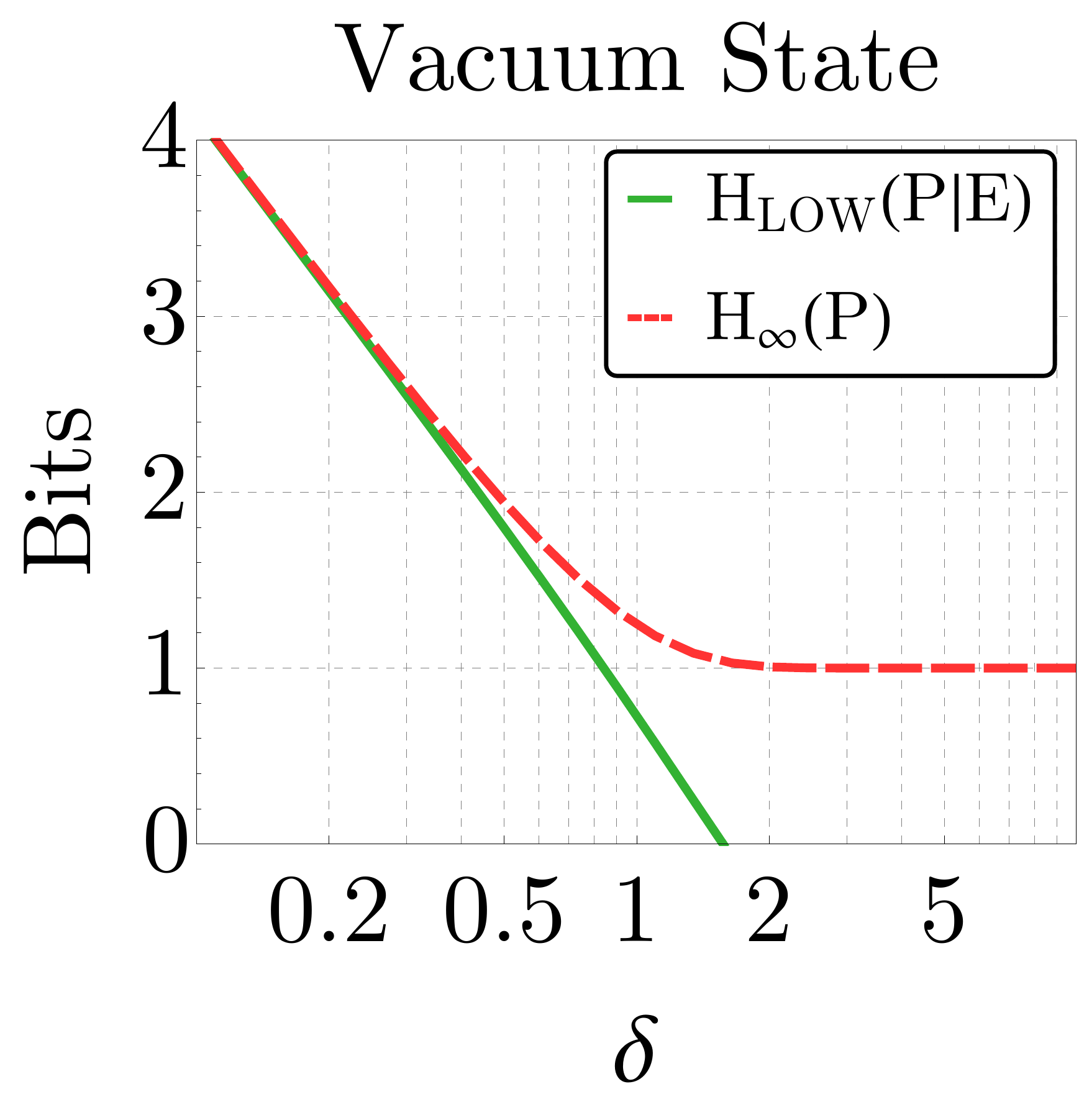}
\includegraphics[width=0.22\textwidth]{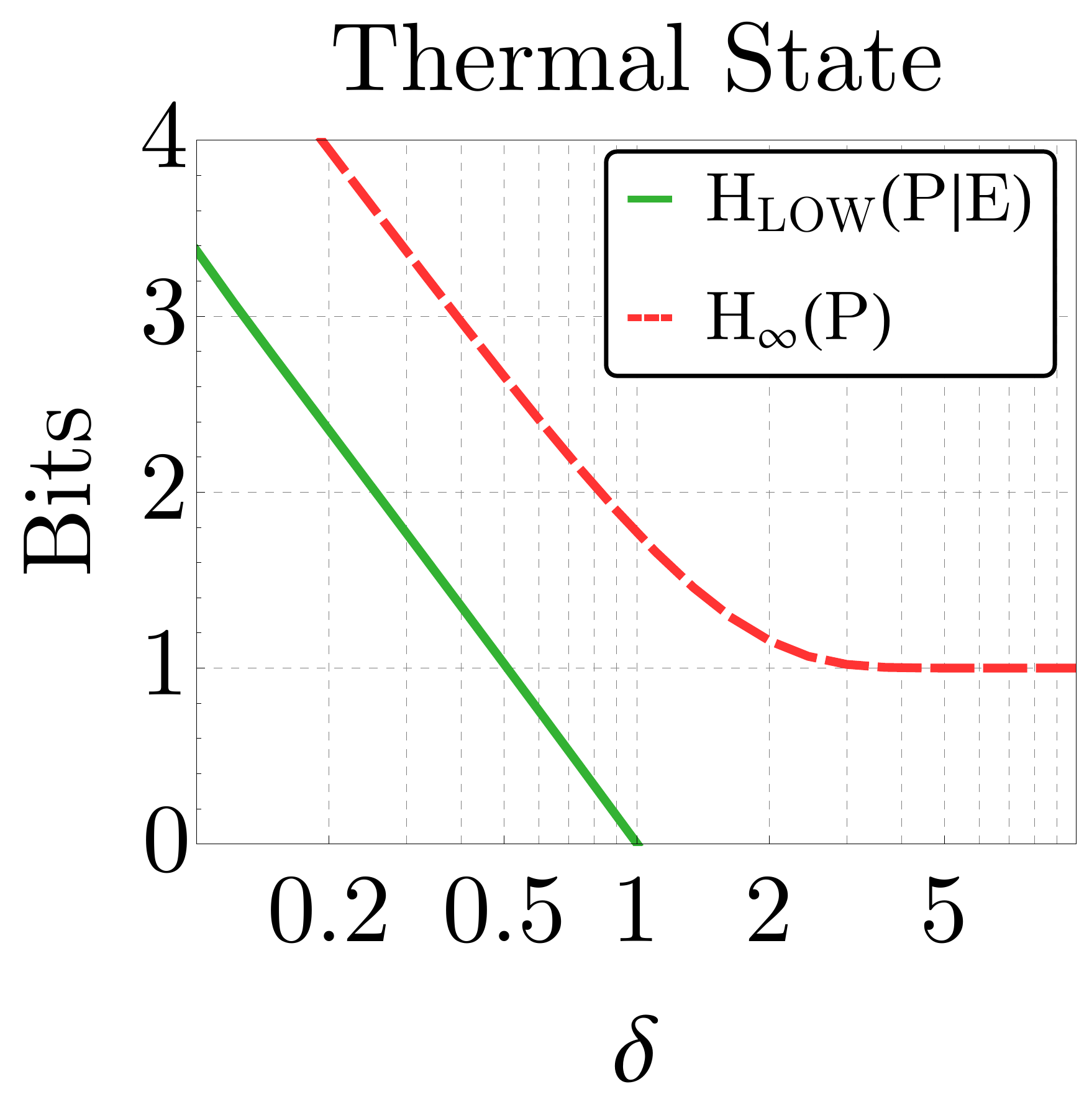}
\caption{Comparison between classical and  conditional min-entropy.  Green solid lines represent the bound on quantum conditional min-entropy, $H_{LOW}(P|E)$, while red dashed lines are the classical min-entropy, $H_{\infty}(P)$, evaluated for a vacuum state (\textbf{left}) and a thermal state  (\textbf{right}) with variance $\sigma^2_{\text{vac}}=1/2$. The conditional min-entropy $H_{\rm min}(P|E)$
lies between $H_{LOW}(P|E)$ and $H_{\infty}(P)$.
For the vacuum state, both estimators attain the same value when the precision of the measurement increases $\delta \rightarrow 0$. For the thermal state, i.e. a mixed state, the classical min-entropy always over-estimates the true content of randomness with respect to the quantum conditional entropy.}
\label{comparison}
\end{figure} 

For a QRNG, the system B is set to trivial and from Eq. \eqref{EUP} it is straightforward to derive
$ H_{\text{min}}(P_{\delta p}|E)\geq H_{\rm LOW}$
with
\begin{equation}
\label{bound}
H_{\rm LOW}(P_{\delta p}|E)\equiv - \log_2 c(\delta q,\delta p)-H_{\text{max}}(Q_{\delta q})\,,
\end{equation}
and $H_{\text{max}}(Q_{\delta q})$ the Renyi entropy of order $1/2$. 

Our method to estimate the content of true random bits for source-device-independent CV-QRNG is summarized in Fig. \ref{setup} and works as follows: I) Alice prepares the state $\rho_A$ (the vacuum or a squeezed vacuum), measures it in the $\hat P$ quadrature (called \emph{data} quadrature) and generates \emph{raw} random numbers; II) the measurement is randomly swapped to the $\hat Q$ quadrature (called \emph{check} quadrature): Alice estimates $H_{\text{max}}(Q_{\delta q})$ by using the outcomes of the check measurements  by 
\begin{equation}
\label{Hmax}
H_{\text{max}}(Q_{\delta q})=2\log_2\sum_k\sqrt{\mathfrak{p}(q_k)}\,;
\end{equation}
III) the bound of $ H_{\text{min}}(P_{\delta p}|E)$ is evaluated  by using Eq. \eqref{bound}; IV) a quantum randomness extractor calibrated on the bound is applied to the raw random numbers. An initial random \emph{seed} for the measurement switching is required, but the protocol is able to quadratically expand the initial randomness as in the protocol introduced in \cite{vallone2014quantum}.

The measurement of the $\hat Q$ operator can be regarded as a tool to estimate, with a partial tomography, whether the state $\rho_A$ is pure or not. In order to better illustrate our approach, in Fig. \ref{comparison} we compare 
the classical min-entropy $ H_{\infty}(P_{\delta p})$ and the bound in Eq. \eqref{bound}  as a function of the precision $\delta\equiv\delta q=\delta p$.
The quantum min-entropy is bounded by these two values, namely 
$H_{LOW}(P_{\delta p}|E)\leq  H_{\text{min}}(P_{\delta p}|E)\leq
H_{\infty}(P_{\delta p})$. 
 Two different input states $\rho_A$ are considered:  the vacuum state, with variance $\sigma^2_{\text{vac}}=1/2$, and a thermal state with variance $\sigma^2_{\text{th}}=1/2+\mu$, where $\mu=2$ is the mean photon number. 
For low $\delta$, the classical min-entropy and the bound can be evaluated analytically, giving $H_\infty(P_\delta)\simeq-\log_2\frac{\delta}{\sqrt{\pi(1+\mu)}}$ and $ H_{\text{LOW}}(P_{\delta p}|E)\simeq H_\infty(P_\delta)-2\log_2\frac{\delta}{\sqrt{2\pi}}\vartheta_3(0,e^{-\delta^2/(2+4\mu)})$ with $\vartheta_3(z,q)\equiv\sum_nq^{n^2}e^{2niz}$ the Jacobi theta-function (see SI). 

For pure states, equality between $ H_{\text{min}}(P_{\delta p})$ and $ H_{\text{min}}(P_{\delta p}|E)$
 is expected and this is the case for the vacuum.
On the other hand, for a thermal state, the classical min-entropy always overestimate the content of true randomness. A thermal state is indeed  mixed and it can be interpreted as $\rho_A$ being  correlated with the environment system $E$: the gap between the two entropies corresponds to the possible leakage of information due to this correlation (Cfr. SI for a detailed discussion).
We also note that when $\delta q$ is large, the bound in Eq. \eqref{bound} underestimates the number of true random bits extractable per measurement because the lower precision implies a looser estimation of the input state. 

\emph{Classical side information -} In our SDI framework, Alice controls and trust the measurement device. 
We assume that Alice optimizes her hardware to not spoil the independence and uniformity of the numbers (e.g. by oversampling the signals, by using unbalanced beam-splitters etc.). We now show that our method take into account effectively also  classical side information. Indeed, even if $\rho_A$ were pure and the generator is optimized, the hardware anyway features an intrinsic classical noise which \virgolette{adds in quadrature} to the quantum signal. 
The result is an increase of the quadrature variance with respect to the shot-noise limit $1/2$, see Si Fig. S3.  For example, for the vacuum input state one observes a variance of $\sigma^{\prime 2}_{\text{vac}}=1/2+\langle n _{\text{noise}}\rangle$
in all quadratures, as for a thermal state. 
Because Alice cannot distinguish whether the input state is mixed or pure, the protocol considers the security most conservative option: any observed \virgolette{mixedness} is treated as if it is caused by some quantum eavesdropping strategy, i.e. the system $A$ entangled with Eve's system $E$. Hence any kind of side information will be erased applying quantum randomness extractors \cite{tomamichel2011leftover,de2012trevisan,renner2011quantum} \emph{properly} calibrated with the conditional min-entropy lower bound. It is clear that the check quadrature has to be measured at random instants. This prevents Eve to carry out deception strategies during the check measurements. Therefore Alice is now able to conservatively bound the amount of true randomness both if an adversary holds a description of the post-measurement classical-quantum state or may get access to the classical noise. It is worth noticing that the real-time estimation of $H_{\text{min}}(P_{\delta p}|E)$ provides a dynamic resiliency against drifts of the classical noise, possibly due to not constant experimental conditions, e.g. temperature variations or interference with external e.m. fields. 

\emph{The experiment -} We built a CV-QRNG based on a homodyne scheme for the measurement of vacuum fluctuations. 
We implemented an all-in-fiber setup with off-the-shelves devices: the local oscillator was a narrow line $1550~nm$ laser connected to one input of a fiber 50:50 beamsplitter and the vacuum entering from the unused port. The exiting ports were connected to a balanced receiver with a bandwidth of 1.6 GHz. The output difference current signal was then sampled at an equivalent rate of 1.25 GS/s by a 12 GHz bandwidth fast oscilloscope (see SI).  We identified the momentum and the position observables as the data and the check quadratures respectively. To simulate the active switching between the complementary observables (which shall be implemented with a LO phase shifter), random subsets were extracted from the whole set of outcomes and attributed to the position quadrature. 
 
As an example, we present the results obtained on a typical run of $m \approx 6 \cdot 10^8$ data samples.  The measured check quadrature variance is $\sigma^{\prime 2}_{\text{vac}}=0.677$, i.e 35.4\% larger than the theoretical value of 1/2, as consequence of the electronic noise which is then treated as an impurity of the input state.  

\begin{figure}[t]
\includegraphics[width=0.42\textwidth]{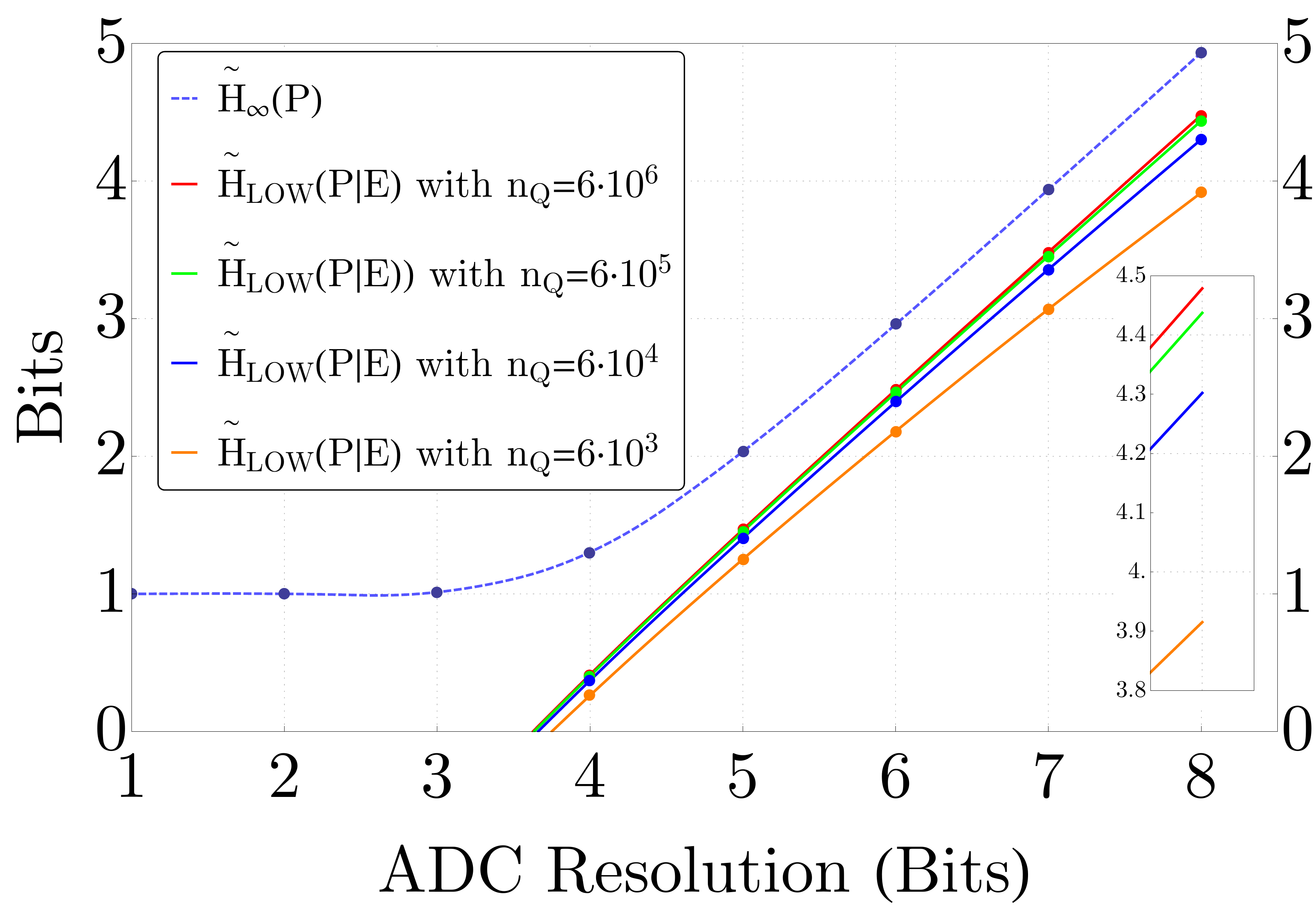}
\caption{Estimated entropy bounds $\widetilde H_{\text{LOW}}(P_{\delta_j}|E)$ for increasing $n_Q$  are reported as function of the measurement precision, i.e. the resolution of the oscilloscope analog to digital converter (ADC). The dashed purple (color online) line corresponds the $H_{\min}(P_{\delta_j})$. The classical min-entropy overestimates the real amount of true random bits extractable, being the measured state equivalent to a thermal state. With a too low precision, the conditional min-entropy becomes not informative because the resolution is too low to distinguish the input state.}
\label{entropie}
\end{figure}

In Fig. \ref{entropie} the lower bounds $\widetilde H_{\text{LOW}}(P_{\delta_j}|E)$ of the entropy are reported as function of the oscilloscope resolution. A $j$-bit resolution corresponds to a precision 
$\delta_j=p_{\rm max}2^{1-j}$ with $p_{\rm max}$ related to the oscilloscope full-scale setting.
We evaluated the bounds for classical systems $P_{\delta_j} $ and $Q_{\delta_j}$ for different resolutions from 1 bit ($\delta_7$) to 8 bits ($\delta_0$). Each point was obtained by averaging 200 bound values $\widetilde H_{\text{LOW}}(P_{\delta_j}|E)=\log_2{\left(\frac{2\pi}{\delta_j^2}S^{00}\left[1,\frac{\delta_j^2}{4}\right]^{-2}\right)}-\widetilde H_{\text{max}}(Q_{\delta_j})$, being $\widetilde H_{\text{max}}(Q_{\delta_j})$ the Bayesan estimator (cfr. \cite{vallone2014quantum}) of the max-entropy evaluated on 200 random subsets of size $n_Q=m\cdot 10^{-i}$ with $i=2 \dots 5$. 

The plot shows the interplay between the resolutions and $n_Q$ in the estimation of the bounds. Indeed, for high precision and large $n_Q$, the reconstruction of the check quadrature distribution bocomes more accurate. In agreement with Fig. \ref{comparison} (Right), classical min-entropy overestimates the real content of entropy. In particular, for $\delta_5$ (resolution of 3 bits) $H_{\infty}(P_{\delta})$ already attains the unit value while the conditional entropy is negative. 
This is because our input state distribution is narrow with respect to the oscilloscope full-range \footnote{in order to limit the probability of having outcomes larger than the full-scale range, see SI}. Therefore the two central bins comprise most of the data points and this prevents Alice to reliably identify the input state, see SI
Fig. S9.

For the generation rate, the number of bits necessary for the switching between the two conjugate quadratures must be accounted: following \cite{vallone2014quantum}, we set $n_Q = \sqrt{m}$ with $m$ the total number of measurements in both the quadratures. Out of the $m$ measurements, the check instants can be chosen in
$\left(
\begin{array}{c}
m \\
n_Q
\end{array}
\right)$ different ways. 
A given random combination then can be encoded in a \emph{seed} $t(m)=\lceil\log_2\frac{m!}{n_Q!(m-n_Q)!}\rceil$ bits long. We evaluated then the secure generation rate, i.e. the net number of true random bits per measurement, according
\begin{equation}
r_{\rm sec}=\frac{1}{m}(m-n_Q)[-c(\delta_j)-\widetilde H_{\max}(Q_{\delta_j})]-t(m)
\end{equation}

Given the total amount of measurements being $m = 615514112$ we employed $n=\lceil{\sqrt{m}}\rceil\simeq 24810$ bits to evaluate the conditional min-entropy. {It is well known that the oscilloscope has an effective resolution lower than the nominal 8 bits, when the sampling rate is high. Hence, we used a conservative bit depth of 5 bits satisfying the SDI requirement of having a trusted and controlled measurement apparatus.} Entropy was estimated to be $\widetilde H_{\text{LOW}}(P_{\delta p}|E)\geq  1.3629$ bits in average per measurement and a corresponding rate of $r = 1.3617$ bits. We generated secure bits at the rate of 1.7 Gbit/s. 

\emph{Conclusions -} At present time, ultimate randomness is reachable only 
by using  device independent protocols of randomness expansion \cite{Pironio2010b} or amplification \cite{Colbeck2012a,Gallego2013}: however, such protocols are highly demanding from an experimental point of view. On the other hand, assuming the absence of local hidden variable theories, true random numbers can be obtained with \virgolette{bottom up} approaches. Methods of recent introduction, evaluate the real content of entropy with an \emph{a priori} characterization of the quantum system: by checking the state purity \cite{Fiorentino2007a, vallone2014quantum} or by checking the quantum system dimensions \cite{lunghi2015self}. Our SDI protocol 
enables the ultra-fast generation of true random numbers. 
Our source-device-independent approach is motivated by  experimental requirements: indeed, it is typically difficult to prepare and keep a real quantum system in a pure state. 
We have showed that unpredictable numbers can be distilled even if the quantum eavesdropper is providing the source of quantum states. 

 Future steps will consider the possibility to merge our protocol with the \virgolette{metrologic} approach introduced by Mitchell et al. \cite{mitchell2015strong} for the analysis of the technical noise added by every component of the hardware. Besides the security advantage, we have demonstrated the feasibility of the protocol with an ultra-fast, cheap and compact CV-QRNG. It is worth to remark that by using commercial balanced receivers and fast LO phase shifter, the secure generation rate can be increased to tens of Gbit/s. Further improvements can envisaged when squeezed states are used as input state for the protocol.

\acknowledgements
\emph{Acknowledgements -} The authors wish to  thank 
R. Corvaja, L. Palmieri and M. Stellini 
for the helpful discussions.
Our work was supported by the Strategic-Research-Project QUINTET of the Department of Information Engineering, University of Padova, the Strategic-Research-Project QUANTUMFUTURE of the University of Padova.

\clearpage
\onecolumngrid

\begin{center}
{\large\bf Appendix: \\
Source-device-independent Ultra-fast Quantum Random Number Generation}
\end{center}

\twocolumngrid


\setcounter{figure}{0}
\renewcommand{\thefigure}{S\arabic{figure}}

\setcounter{equation}{0}
\renewcommand{\theequation}{S\arabic{equation}}

\section{Discretization in CV-QRNG}

CV-QRNGs are based on the measurement of quantum fluctuations  of an electromagnetic (e.m.) mode quadrature by means of \emph{optical homodyne}: a  quantum state $\omega_A$ of a e.m. mode is combined with a classical field, the so called local oscillator (LO), by a beam splitter.
  
For a given mode of the electromagnetic field, the generic quadrature operator can be expressed by $\hat{q}(\varphi) = 2^{-\frac{1}{2}}\,\left( e^{i\frac\varphi2} \hat{a}^\dag+e^{-i\frac\varphi2} \hat{a} \right)$ being  $\hat{a}^{\dag}$ and $\hat{a}$ the creation and annihilation operators such that $[\hat{a},\hat{a}^\dag]=1$ holds. The canonically conjugated  operators usually  identified as the \emph{position} $\mathcal{Q}$ and \emph{momentum} $\mathcal{P}$  quadratures observables are given by $\hat{Q}\equiv\hat{q}(0) $ and $\hat{P}\equiv\hat{q}(\pi)$. They satisfy $[\hat{Q},\hat{P}]=i$. The eigenvalues equations for position and momentum operators are given by $\hat{Q}\ket{q} =q \ket{q}$  with $q\in\mathbb R$ and $\hat{P}\ket{p} =p \ket{p}$ with $p\in \mathbb R$.

If  $\rho_A$ is assumed to be the vacuum state, i.e. $\rho_A=\ket{0}\bra{0}$, its position and momentum representation are respectively given by $\psi_0(q)\equiv\braket{q}{0}=\pi^{-\frac{1}{4}}e^{-q^2/2}$ and $\widetilde\psi_0(p) \equiv \braket{p}{0}=\pi^{-\frac{1}{4}}e^{-p^2/2}$.
For the vacuum $\ket{0}$, the probability distributions $Q(q)\equiv |\psi_0(q)|^2$ and $P(p)\equiv |\widetilde \psi_0(p)|^2$ are Gaussian distribution with null average value and variance $\sigma_{\text{vac}}^2=\nicefrac{1}{2}$.

 With CV systems, the unavoidable discretization of the measurements due to the finite resolution of the experimental devices has to be considered. More specifically, a coarse grained version of operators 
 can be obtained by introducing a partition $\mathcal P_{\delta p}=\{I_{\delta p}^k \}_{k=-\infty}^{+\infty}$  of the measure space   $\mathbb{R}$ \cite{Furrer2014}. The elements $I_{\delta p}^k $ are given by half-open intervals  such that  $I_{\delta p}^k =\left(k\delta p,(k+1)\delta p\right]$ where $\delta p$ is the \emph{precision} of the measurement and $k\in \mathbb{N}$.
Alice applies  POVMs $\left\{\hat{P}^k_{\delta p}\right\}$ with elements $\hat{P}^k_{\delta p}=\int_{k\delta p}^{(k+1)\delta p}dp\ket{p}\bra{p}$ on $A$ and she stores the outcomes $p_k$ in the classical  system (register) $P_{\delta p}$. The post-measurement state of $P_{\delta p}$ corresponds to the probability distribution of $p_k$, and it is given by $\rho_P=\sum_k \mathfrak{p}(p_k){\hat{P}^k_{\delta p}}$ 
where  $ \mathfrak{p}(p_k)=\text{Tr}\left[\omega_A\hat{P}^k_{\delta p}\right]=\int_k^{(k+1)\delta p}dp\bra{p}{\rho_A}\ket{p}$. 
Similarly, the discretized $\hat Q$
operator is given by the POVMs
$\left\{\hat{Q}^k_{\delta q}\right\}$ with elements $\hat{Q}^k_{\delta q}=\int_{k\delta q}^{(k+1)\delta q}dq\ket{q}\bra{q}$.

The estimation of max-entropy $H_{\rm max}(Q_{\delta q})$ is based on the relative frequency of the outcomes of the discretized $\hat Q$ operator, as given by \eqref{Hmax}:
\begin{equation}
\label{HmaxSI}
2^{H_{\rm max}(Q_{\delta q})}=
{\left(\sum^{+\infty}_{k=-\infty}\sqrt{\mathfrak{p}(q_k)}\right)}^2
\end{equation}
While the sum in the above equation extends from $-\infty$ to $+\infty$, experimental outcomes range from $-M$ to $+M$ due to experimental finite measurement range. Outcomes which exceed this range 
are registered as $M+1$ or $-M-1$ outcomes.  
We estimated the max-entropy by limiting the sum in 
\eqref{HmaxSI} from $-M$ to $M$ and
by neglecting the term $
\sum_{k>|M|}\sqrt{\mathfrak{p}{(p_k)}}$.
We can upper bound such neglected terms, by
considering a trial with a total of $N$ measurements: defining 
$P_M= \sum_{k=-\infty}^{-M}\mathfrak{p}{(p_k)}+\sum_{k=M}^{\infty}\mathfrak{p}{(p_k)}$,
we expect that $n\sim P_MN$ events result in an outcome out of range. 
The worst scenario, that maximize the
neglected term, is given when each of the  $n$ outcomes falls into a different bin. In this situation, we have  $\mathfrak{p}(p_k) \approx \nicefrac{1}{N}$ such that $\sum_{k>|M|}\sqrt{\mathfrak{p}{(p_k)}}\leq \frac{n}{\sqrt{N}}\approx \sqrt{N}P_M$.
Since $P_M$ corresponds to the double sided tail probability  of the Gaussian distribution, a narrow distribution (i.e. a small standard deviation $\sigma$ compared to $M\delta$) corresponds to a low error in the min-entropy estimation.  For the experimental data presented in the main text, we have that $M\delta\approx 10.5 \sigma$, with the max entropy evaluated on $N\approx 25\cdot 10^3$ measurements.  Then, we estimate that the error introduced by the finite measurement range, is of order of $10^{-26}$.

\section{Example of the difference between the classical $H_\infty$ and the quantum min-entropy $H_{\rm min}$}
In this section we show the difference between the classical and quantum min-entropy for a CV-QRNG. In particular, we will compare QRNGs based on two different inputs, a squeezed-state and a thermal state with the same output momentum distribution. We will show, that by measuring the system only in the momentum quadrature, it is not possible to correctly evaluate the conditional quantum min-entropy in both cases.

Let's first consider a CV-QRNG generator based on a $\zeta$-squeezed state as input. The wavefunction of a squeezed state is written in the $\mathcal Q$ and $\mathcal P$ space respectively as $\psi_{\zeta}(q)=\sqrt{\frac{\zeta}{\pi^{\nicefrac{1}{2}}}}e^{\nicefrac{-(\zeta q)^2}{2}}$ or $\widetilde\psi_{\zeta}(p)=\frac{1}{\sqrt{\zeta\pi^{\nicefrac{1}{2}}}}e^{\nicefrac{-p^2}{2\zeta^2}}$. When $\zeta>1$, Alice generates random numbers by measuring the momentum quadrature $\mathcal{P}$ because its variance $\sigma^2_\mathcal{P}\equiv\zeta^2/2$ is larger than the position variance $\sigma^2_\mathcal{Q}\equiv1/(2\zeta^2)$.

\begin{figure}[h!]
\begin{center}
\includegraphics[width=0.98\linewidth]{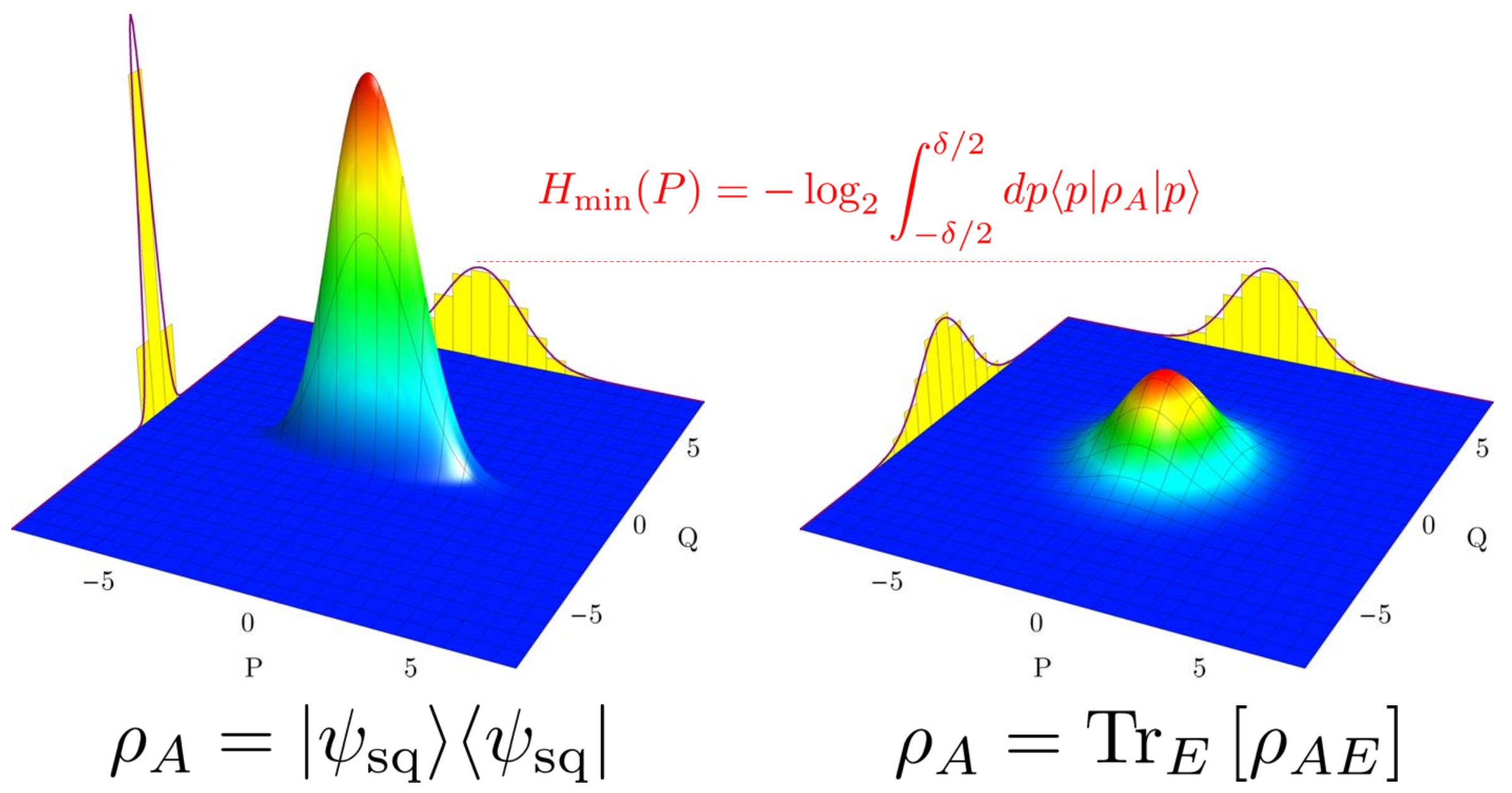}
\end{center}
\caption{\textbf{left}: Wigner function for a  $\mathcal{Q}$-squeezed vacuum state ($\zeta=2$) and the relative discretized probability distribution (yellow histograms) for the two conjugate quadratures. Since the outcome distribution for  $\mathcal{P}$ is wider, the outcomes of momentum measurements (performed with precision $\delta p$) are used as random numbers.  This is an ideal input state for a CV-QRNG: the state is pure and the randomness extractor can be calibrated by the classical min-entropy $H_{\min}(P_{\delta p})$. \textbf{right}: Wigner function
of a thermal state, that can be purified by a two-mode squeezed
vacuum. The probability distribution for the $P$ outcomes coincides with the 
distribution obtained with the $\mathcal{Q}$-squeezed vacuum state.
In this case, the classical min-entropy over-estimates 
the true content of randomness, because it does not take into account
the quantum side information possessed by Eve.
}
\label{figSI:example}
\end{figure}

The outcomes of the momentum quadrature 
follow a Gaussian distribution given by
\begin{equation}
\label{squeezedP}
\mathfrak{p}(p_k)=
{\rm Tr}_A[\hat P^k_{\delta p}\rho_{A}]=
\frac{1}{	\sqrt{\pi\zeta^2}}\int^{(k+1/2)\delta p}_{(k-1/2)\delta p}
e^{\nicefrac{-p^2}{\zeta^2}}{\rm d}p .
\end{equation}
As said in the main text, 
the amount of true random bits can be evaluated by the
quantum conditional min-entropy.
In this case, since the input state is pure, 
the quantum min-entropy is equal to the classical min-entropy  evaluated on the outcomes.
\begin{equation}
\begin{aligned}
H_{\infty}(P_{\delta p})=&-\log_2[\max_k \mathfrak{p}(p_k)]
\\
=&-\log_2\frac{1}{	\sqrt{\pi\zeta^2}}\int^{\delta p/2}_{-\delta p/2}
e^{\nicefrac{-p^2}{\zeta^2}}{\rm d}p 
\end{aligned}
\end{equation}
Indeed, since the state is pure, 
the optimal eavesdropper strategy is to bet on the most likely results.

The same output momentum quadrature distribution \eqref{squeezedP} can be obtained when the input state is thermal {$\rho_A=\frac{1}{\cosh^2\xi}\sum_n(\tanh\xi)^{2n}\ket{n}\bra{n}$
and $2\sinh^2\xi=\zeta^2-1$ (see Fig. \ref{figSI:example}). The mean photon number
of such thermal state is given by $\mu=\sinh^2\xi$.}
The mixedness of the thermal state implies a 
correlation with the environment.
For instance, a purification $\rho_{AE}$ of $\rho_A$ is represented by the 
\emph{two mode squeezed vacuum state},
$\ket{\Psi_\xi}=\frac{1}{\cosh\xi}\sum_n(\tanh\xi)^n\ket{n}_A\ket{n}_E$, where $\xi$ is a squeezing parameter.
This state can be regarded as an optical approximate version of an EPR entangled state and it has been recently used in a CV-QKD scheme \cite{Eberle2013a}. 
If Eve controls the system $E$, she can 
gain information on the quadrature outputs
measured by Alice. Indeed, 
measurements performed on each system of the entangled pair give (anti-) correlated outputs with high probabilities.

The conditional min-entropy is now lower that the classical min-entropy and
its value should be evaluated by the following equation:
\begin{equation}
H_{\rm min}(P_{\delta p}|E)=-\log_2\max_{\{\mathcal E_k\}}
\sum_k\mathfrak{p}(p_k){\rm Tr}[\mathcal E_k\rho^{(E)}_k]\,.
\end{equation}
In the previous equation $\rho^{(E)}_k$ is the quantum state hold by Eve when the
output $k$ is obtained, namely:
\begin{equation}
\rho^{(E)}_k={\rm Tr}_A[\hat P^k_{\delta p}\rho_{AE}]\,,
\end{equation}
while the maximization is performed over general POVM $\{\mathcal E_k\}$ that
Eve could measure.
The conditional min-entropy is hard to evaluate,
but a lower bound can be obtained by the measurement in the conjugate quadrature $Q$, as explained in the main text:
\begin{equation}
\label{HboundSI}
H_{\rm min}(P_{\delta p}|E)\geq
H_{\rm LOW}(P_{\delta p}|E)
\end{equation}
with
\begin{equation}
\label{HlowSI}H_{\rm LOW}(P_{\delta p}|E)\equiv
- \log_2 c(\delta q,\delta p)-2\log_2\sum_k\sqrt{\mathfrak{p}(q_k)}
\end{equation}

\begin{figure}[t!]
\centering
\includegraphics[width=0.3\textwidth]{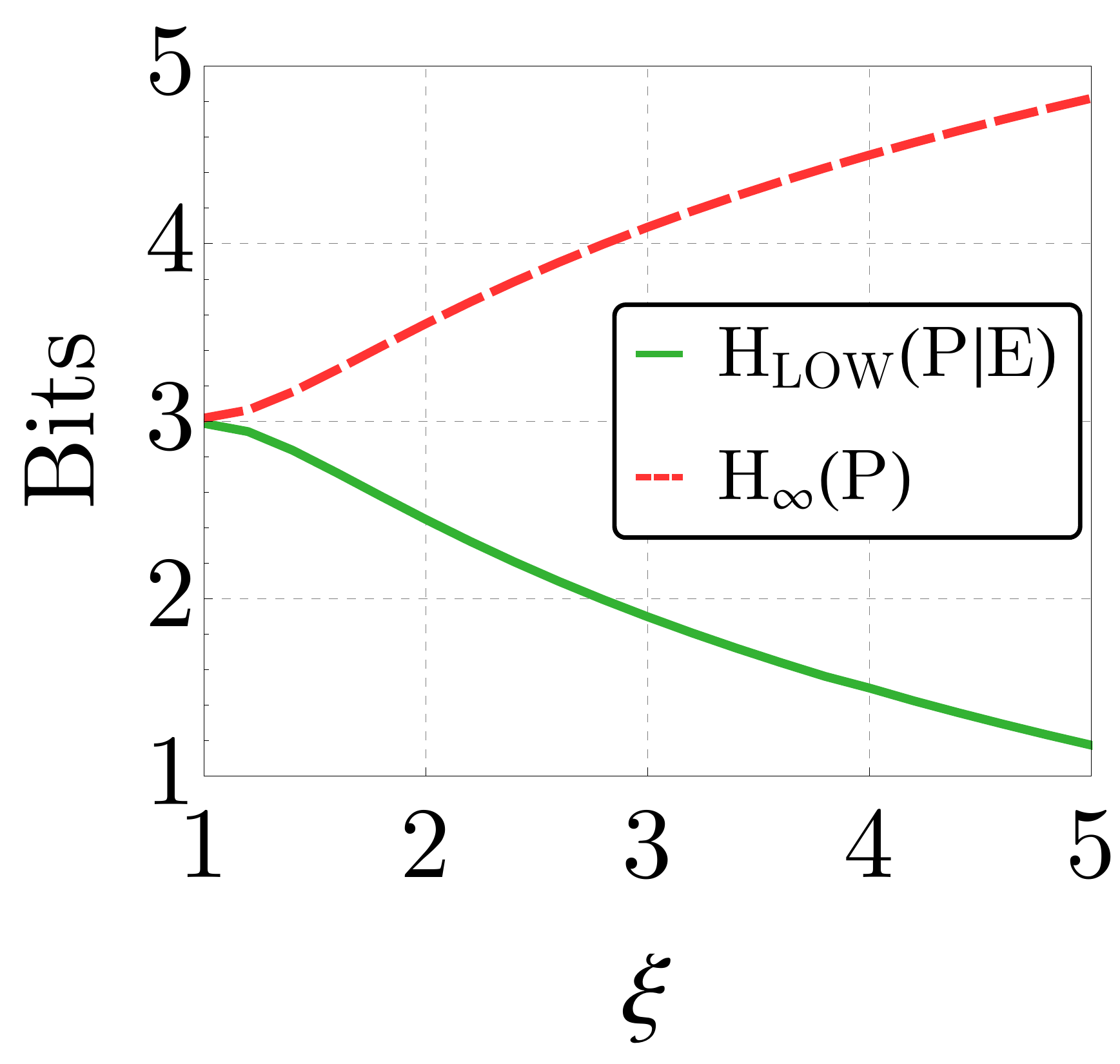}
\caption{\textbf{Left}: $H_{\text{min}}(P)$ (red dashed line) and $H_{\text{LOW}}(P|E)$ (green line) are reported as function of the degree of squeezing $\zeta$ for a given precision ($\delta = 0.22$).}
\label{figSI:twinentropy}
\end{figure}

In Figure \ref{figSI:example} { we show the Wigner functions of the squeezed and thermal state with
the same $\mathcal P$ distribution.} The difference between the two $\mathcal Q$ distribution is evident. 
The impurity of the thermal state may be indeed detected by Alice by observing that that the (check) $\mathcal Q$ quadrature variance is not squeezed.
However, it is not necessary to abort the protocol: Alice can still extract random bits by calibrating a quantum randomness extractor with the conditional min-entropy. The estimation of a lower, conservative, bound to this quantity can be obtained by the relation \eqref{HboundSI}.

In Figure \ref{figSI:twinentropy}, we show the value of the classical min-entropy and the lower bound  $H_{\text{LOW}}(P|E)$ for a thermal state. 
It is worth noticing that
high squeezing corresponds to strong correlations 
with the Environment: by increasing $\zeta$, Eve may gain more information about Alice system:  $H_{\text{min}}(P_{\delta p}|E)$ decreases because the fraction of bits known by Eve becomes larger.
On the other side, $H_{\min}(P_{\delta p})$  increases because the distribution of momentum outcomes 
approaches to the uniform distribution.

This examples stresses the complete unsuitability of the use of classical min-entropy as an estimator for the security of QRNG: the application of a randomness extractor based on this estimator does not allow the user Alice to eliminate the full quantum side information possessed by Eve. 

The quantities $H_{\text{LOW}}(P|E)$ and
$H_{\infty}(P)$ can be evaluated analytically in the approximation
of high precision (namely for $\delta\rightarrow 0$).
For a thermal state $\rho_{\text{Th}}$ with average photon number $\mu$ 
the probability to obtain a given outcome $q_k$ or $p_k$ is given,
for small $\delta q$ and $\delta p$, by
\begin{equation}
 \mathfrak{p}(q_k)\simeq  \frac{\delta q\,e^{-\frac{(\delta q\, k)^2}{1+2\mu}}}{\sqrt{\pi(1+2\mu)}} \,,
\qquad
\mathfrak{p}(p_k)\simeq  \frac{\delta p\,e^{-\frac{(\delta p\, k)^2}{1+2\mu}}}{\sqrt{\pi(1+2\mu)}} \,\,.
\end{equation}
The max-entropy for the $\mathcal{Q}$ quadrature can be written as $H_{\max}(Q_{\delta q})=2\log_2\sum_k\sqrt{ \mathfrak{p}(q_k)}$,
an it can be explicitly evaluated as:
\begin{equation}
\begin{aligned}
H_{\max}(Q_{\delta q})
&= \log_2 \frac{\delta q}{\sqrt{\pi(1+2\mu)}}+2\log_2\sum_k e^{-\frac{(\delta q k)^2}{2(1+2\mu)}} \,\,.
\end{aligned}
\end{equation}
On the other side, the classical min-entropy is given by
\begin{equation}
H_{P_\delta}\equiv -\max \log_2(\mathfrak{p}(p_k))\simeq
-\log_2\frac{\delta p}{\sqrt{\pi(1+2\mu)}}
\end{equation}

Considering small $\delta q$ and $\delta p$ and using the above max-entropy,
the bound on the conditional min-entropy (eq. \eqref{bound} of the main
text) becomes:
\begin{equation}
\begin{aligned}
H_{LOW}(P_{\delta p}|E) 
&=H_{\min}(P_{\delta p})-2\log_2\frac{\delta q}{\sqrt{2\pi}}
\vartheta_3(0,e^{-\frac{(\delta q)^2}{2(1+2\mu)}})\,\,.
\end{aligned}
\end{equation}
with
$\vartheta_3(z,q)$ the Jacobi theta-function. Since
$x\,\vartheta_3(0,e^{-x^2})\geq \sqrt{\pi}$ for $x\geq 0$ it is easy to prove that the r.h.s. is always lower than the classical min-entropy
$H_{\min}(P_{\delta p})$.
Moreover, since $\lim_{x\rightarrow 0}x\vartheta_3(0,e^{-x^2})=\sqrt{\pi}$,
in the limit of infinite precision $\delta q,\delta p\rightarrow 0$ we have
\begin{equation}
H_{LOW}(P_{\delta p}|E) 
\sim H_{\min}(P_{\delta p})-\log_2(1+2\mu)
\end{equation}

For the pure vacuum state, corresponding to $\mu=0$ the bound coincides with 
classical min-entropy (see Fig. 1 of the main text for $\delta\rightarrow0$).
For a thermal state with $\mu>0$, the additional term $\log_2(1+2\mu)$ prevents to obtain the equality between the two entropies:  also in case of infinite precision the bound of the conditional min-entropy is always lower than the classical min-entropy.

\begin{figure}[t]
\begin{center}
\includegraphics[width=\linewidth]{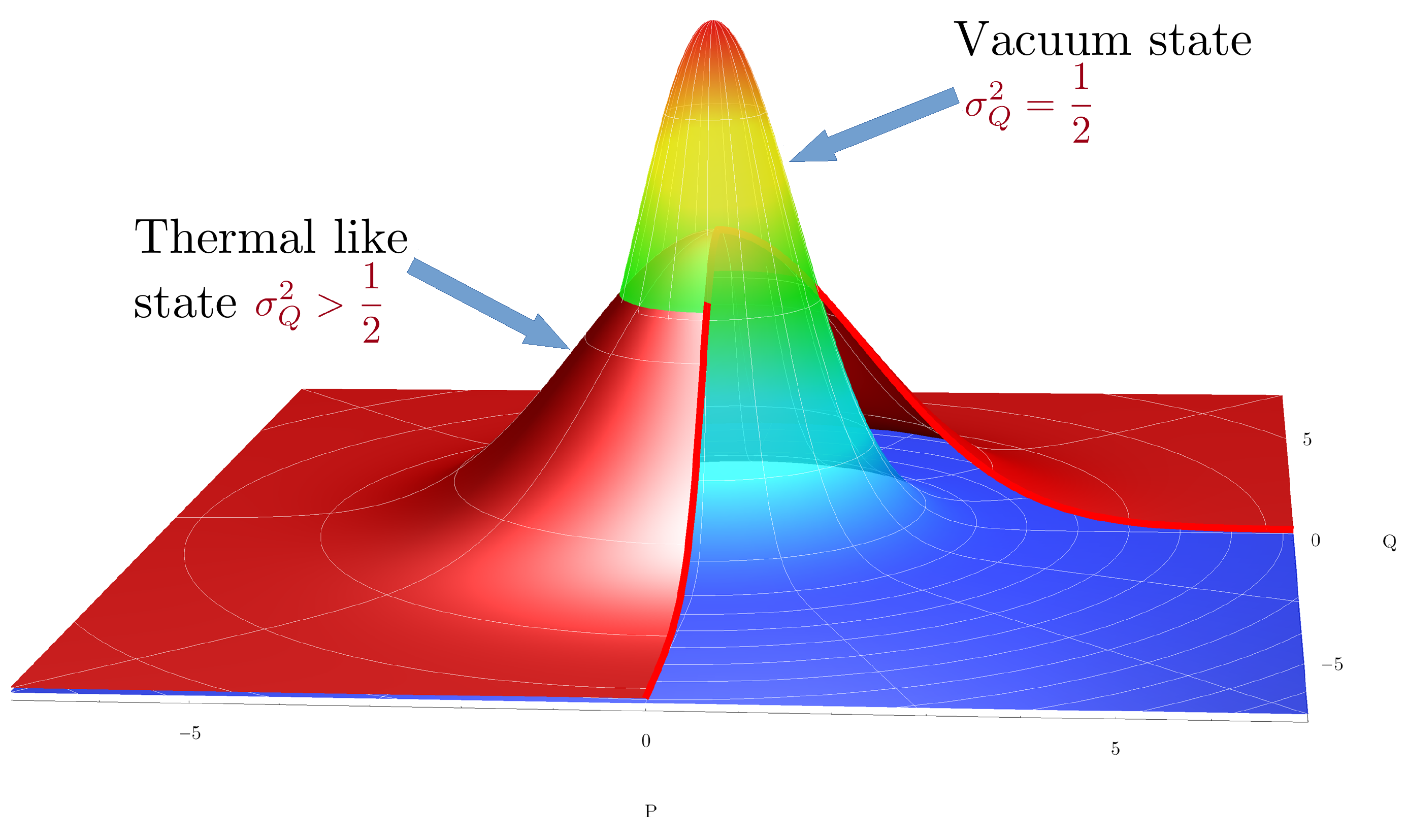}
\end{center}
\caption{The classical noise of the generator components cannot be separated from the quantum noise of the vacuum fluctuations: the neat effect therefore is a widening of the measured vacuum state outcome distribution as if the input state was a thermal state.}
\label{mexicans}
\end{figure}

\section{Experimental Setup}
\begin{figure}[t]
\begin{center}
\includegraphics[width=\linewidth]{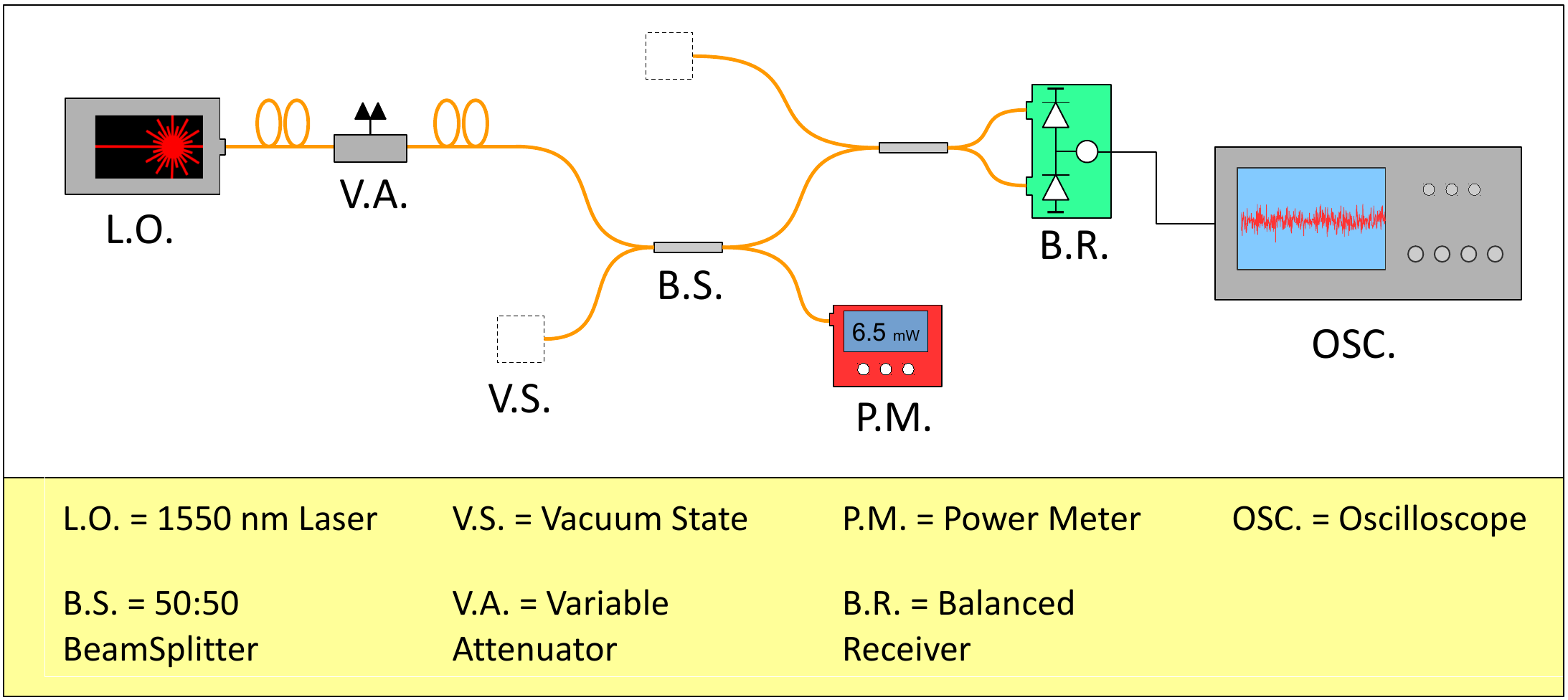}
\end{center}
\caption{In Figure a scheme of the experimental setup is reported. The QRNG was realized with fiber coupled commercial components.}
\label{figSI:setup}
\end{figure}

A scheme of the experimental setup is reported in Fig. \ref{figSI:setup}. The local oscillator is provided by a \textbf{Thorlabs SFL-1550} fiber coupled laser 
centered at 1550 nm. The laser is driven by a current and temperature controller which keep the laser operating in a single mode region. The laser output is connected to a \textbf{Thorlabs VOA50} single mode fiber broadband variable attenuator. This device is essential to control the local oscillator power without modifying the optimal working region. The attenuator is connected to one input of a first 50:50 fiber beamsplitter, with the two outputs which go to a power meter and to a second 50:50 fiber beamsplitter, respectively. The power meter is used to monitor the power, while the last beamsplitter brings the signal into the balanced receiver. The superposition of the LO and the vacuum states entering from the unused ports of the beamsplitters, is then detected and converted in current by a pair of InGaAs PINs included in the single self-contained \textbf{Thorlabs PDB480C}. This device, with a nominal bandwidth of 1.6 GHz,  takes the difference and amplifies the PIN signals. Such monolithic configuration helps consistently to reduce the coupling with environmental electromagnetic noise. In addition to a low level of intrinsic technical noise (combination of thermal, dark current, amplifiers noise), the device features a common mode rejection ratio (CMRR) of over $30 \,dB$ in line with the performances of other receivers used in Literature for CV-QRNG. These two characteristics are of main relevance. On one hand, intrinsic technical noise does not cancel by taking the difference of the signals: then, a lower detector noise corresponds to an higher quantum-signal to classical-noise ratio. On the other hand, the more the PIN are matched in responsivity, the more effective is the cancellation of the external classical noise affecting both the modes (e.g. spurious oscillation of the LO). It is worth to remark that the setup components are \virgolette{commercial of the shelves}, (COTS). The use of COTS devices was motivated by the possibility to demonstrate the feasibility of the method and how security can be provided to CV-QRNG for the common use. 

The final stage of the setup consisted of an oscilloscope \textbf{Tektronix TDS6124C} featuring a bandwidth of 12 GHz was used as ADC. The oscilloscope was remotely controlled with a personal computer for the logging of the waveforms which later analyzed to get the raw random numbers and the check basis numbers to apply the EUP protocol.

\begin{figure}[ht]
\begin{center}
\includegraphics[width=\linewidth]{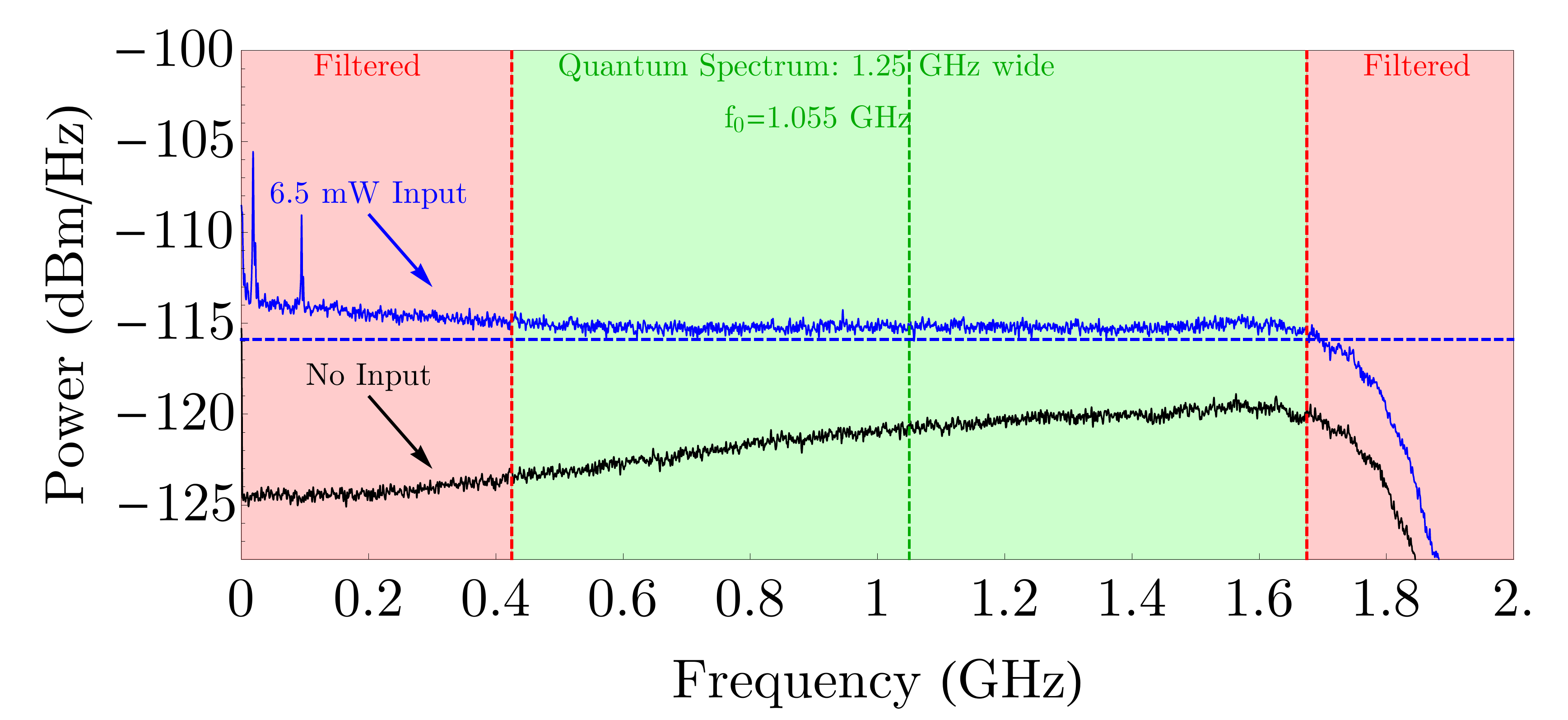}
\end{center}
\caption{The power spectral density function of the signals with LO turned off (black line) and with a LO power of 6 mW (blue line) is reported. The green shaded region identifies the 1.250 GHz wide region of the spectrum which was considered for the extraction of the raw random numbers.  The signal has been downmixed with a sinusoidal carrier at frequency $f_0 = 1.055$ GHz and then filtered with a low-pass filter with 625 MHz cut off frequency.}
\label{figSI:spectrum_CV}
\end{figure}

In Fig. \ref{figSI:spectrum_CV}, the typical Power Spectral Density (PSD) of the output signal is reported.  In particular the PSD with the LO turned off  (black trace) and  with a 6.5 mW LO (blue trace) is reported. In order to filter out those regions of the spectrum affected by technical noise and to enhance the signal-to-noise ratio, we digitally downmixed and low-pass filtered the signal. We considered a flat region 1.250 GHz wide, with optimal central frequency at $f_0 = 1.055$ GHz. 

\begin{figure}[ht]
\centering
\includegraphics[width=\linewidth]{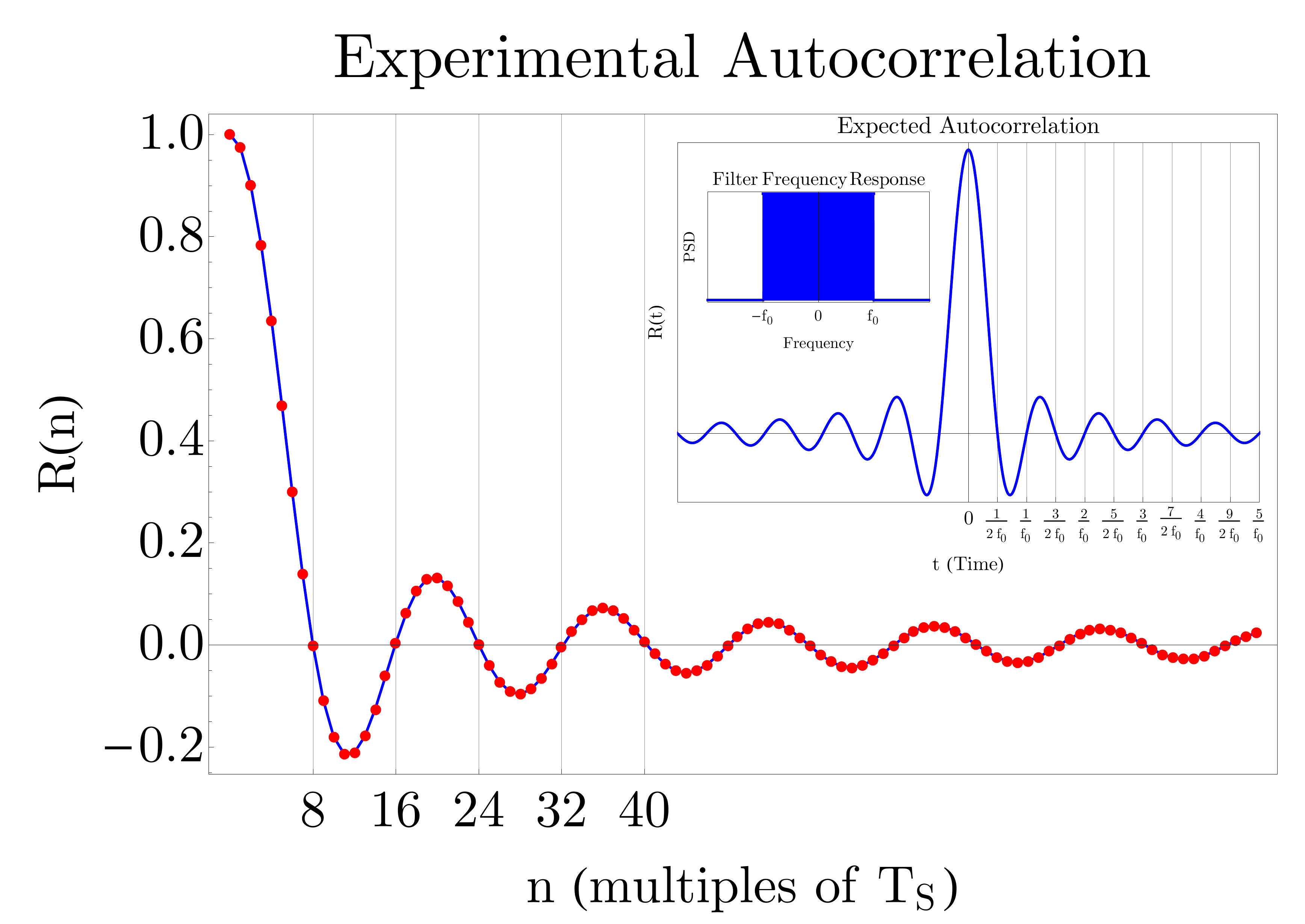}
\includegraphics[width=\linewidth]{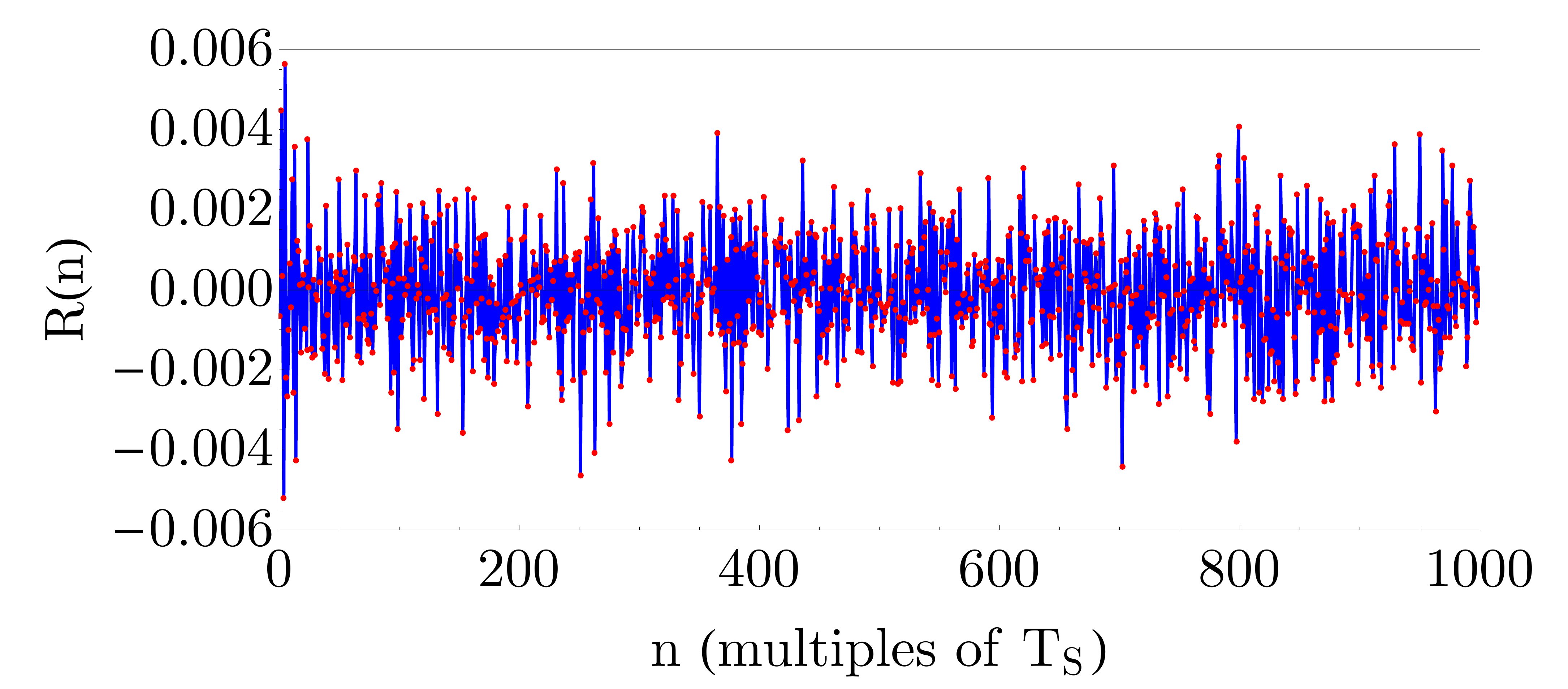}
\caption{\textbf{Top}: experimental autocorrelation of the filtered data as function of the temporal separation in multiple of the sampling interval $T_S$. The correlation is modulated according a sinc function. This is indeed the expected behavior once that a signal is filtered by a low pass filter, top inset. By means of the Wiener-Kitchine theorem one can analytically calculate the zeros of the autocorrelation and then the corresponding down sampling frequency in order to achieve a null self-correlation. \textbf{Bottom:} by downsampling the original waveforms, the quadrature measurements become uncorrelated.}
\label{figSI:corre}
\end{figure}
\begin{figure}[ht]
\centering
\includegraphics[width=\linewidth]{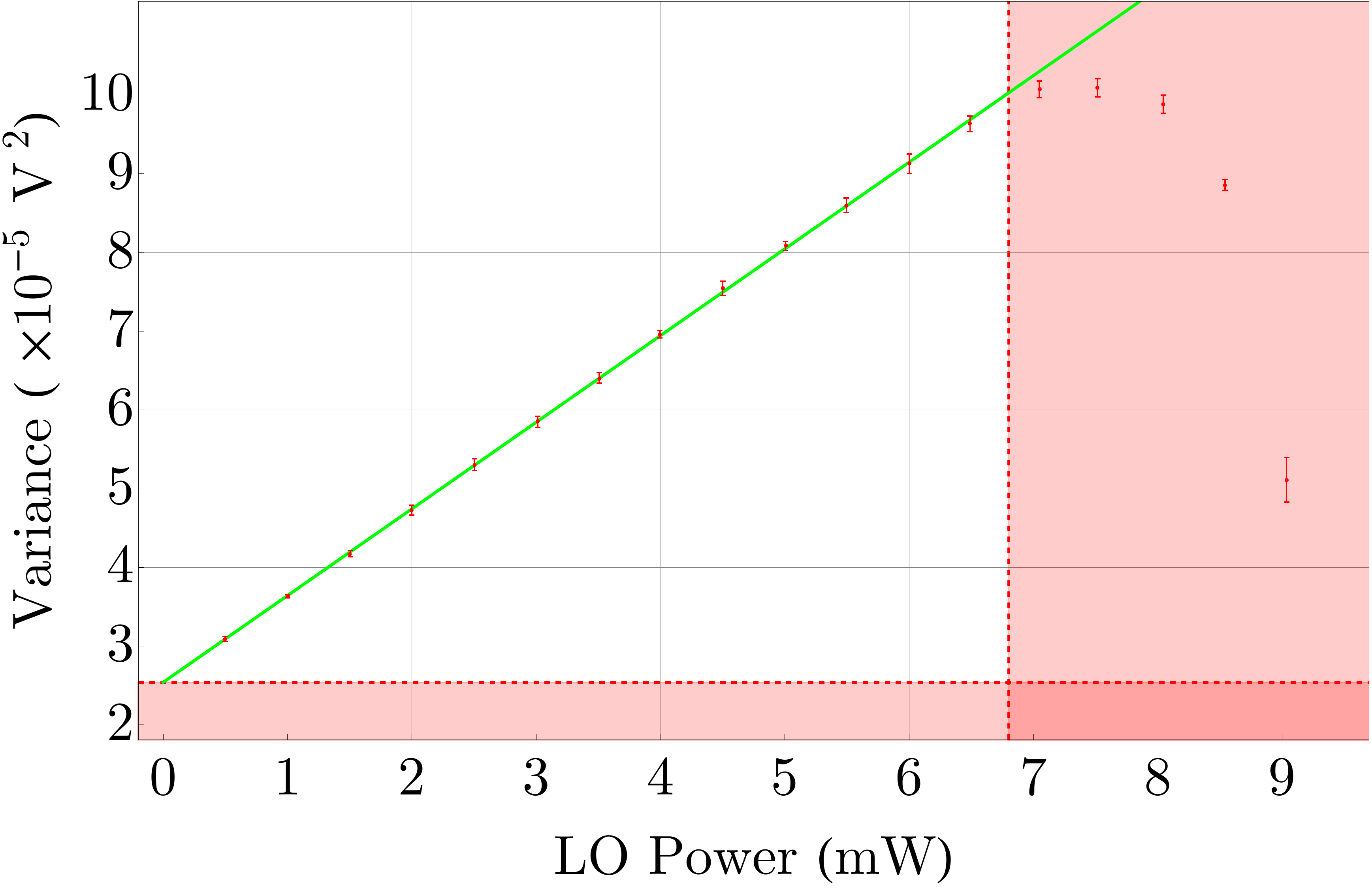}
\caption{The plot shows the linear relation between the LO power and the measured voltages variances. Between 0.5 mW and 7 mW we are in a QNL region whereas at 7.5 mW transimpendance amplifiers start to saturate with a not linear response of the detectors, evidenced in the red shaded region on the right. The red shaded region below represents the contribution of the technical noise.}
\label{figSI:rel_linear}
\end{figure}

The \emph{quadrature signal} was obtained by digital performing a downsampling to 1.25 GSamples/s, in order to match the Nyquist frequency of the low-pass filter and to eliminate the correlations due to  oversampling. On this regard in  Fig. \ref{figSI:corre}, we report  the correlation function before (top) and after (down) the downsampling. Before downsampling, the correlation shows the sinc modulation imposed by the low-pass filter. After downsampling, the residual correlation is, in average, of two orders of magnitude lower. 

In Figure \ref{figSI:rel_linear} we show the variance $\sigma^2$ of \emph{quadrature signal} in function of the LO power.  We varied the LO power between 0.5 mW and 9 mW in steps of 0.5 mW. The linearity is clearly evident between 0.5 and 6.6 mW. Above 7.0 mW the loss of linearity can be ascribed to PIN and transimpendance amplifiers saturation, cfr. \cite{bachor2004guide} and \cite{gray1998photodetector}.

By considering the region between 0.5 mW and 6.5 mW, a linear fit gives the angular coefficient $m= 0.0108 \,\, {V^2}/{W}$  and the intercept $a = 2.579 \cdot 10^{-5} \,\, V^2$. The presence of a not-null intercept $a\neq 0$ is the signature of the experimental noise having a non-quantum origin. At the base of the plot, the red shaded area marks the region where technical noise is dominant. This noise can not be eliminated and it ends up in the signal generating the raw random numbers, affecting their security.

\begin{figure}[h]
\includegraphics[width=0.9\linewidth]{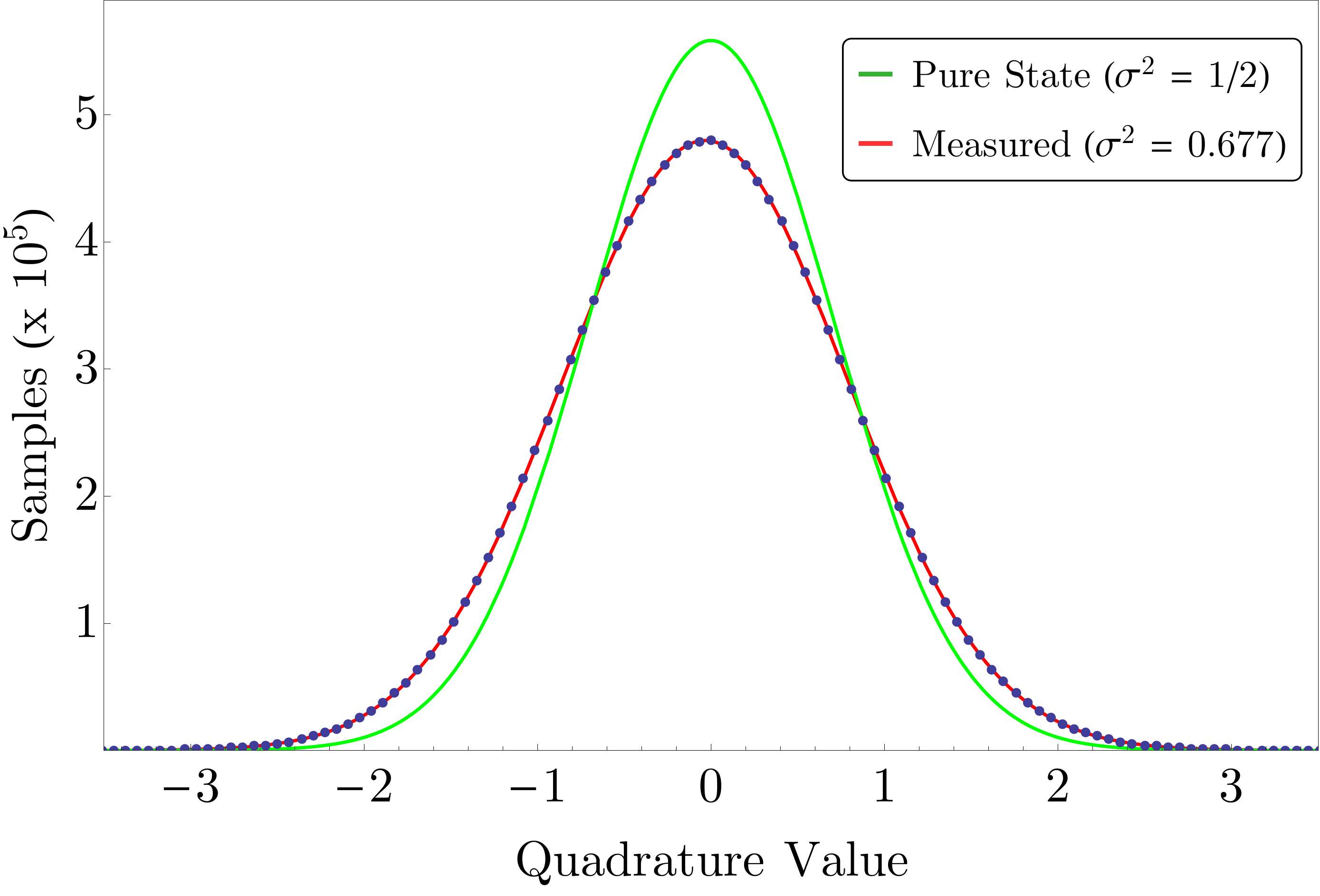}
\caption{In Figure, the comparison between the ideal probability distribution of a pure state (green solid line) and the fitted distribution (red solid line) of the actual experiment data (blue points).)}
\label{SI:quad_ditr}
\end{figure}

As reported in the Main Text, we consider a typical run of $m \approx 6 \cdot 10^8$ data samples. In Fig. \ref{SI:quad_ditr} the expected quadrature distribution (green solid line) is compared with the distribution obtained by fitting the experimental data (red solid line and blue points respectively). A widening of the quadrature variance is observed, $\sigma^{\prime 2}_{\text{vac}}=0.677$ instead of the ideal $\sigma^{2}_{\text{vac}}=1/2$, see Fig. \ref{SI:quad_ditr}. Because is it not possible to discriminate whether the source of the extra noise resides in the preparation or in the measurement stage, according to our protocol the input state is treated as mixed state detected by a perfect detector.  As specified in the Main Text, to simulate the estimation the entropy bounds with the active switching of the measurement bases, we extracted random subsets of samples that were attributed to the as \virgolette{check position quadrature}. In the following, the amount of secure random bits that can be extracted in function of the ADC precision and of number of check measurements, will be presented.

\begin{figure}[h!]
\begin{center}
\includegraphics[height=0.95\linewidth]{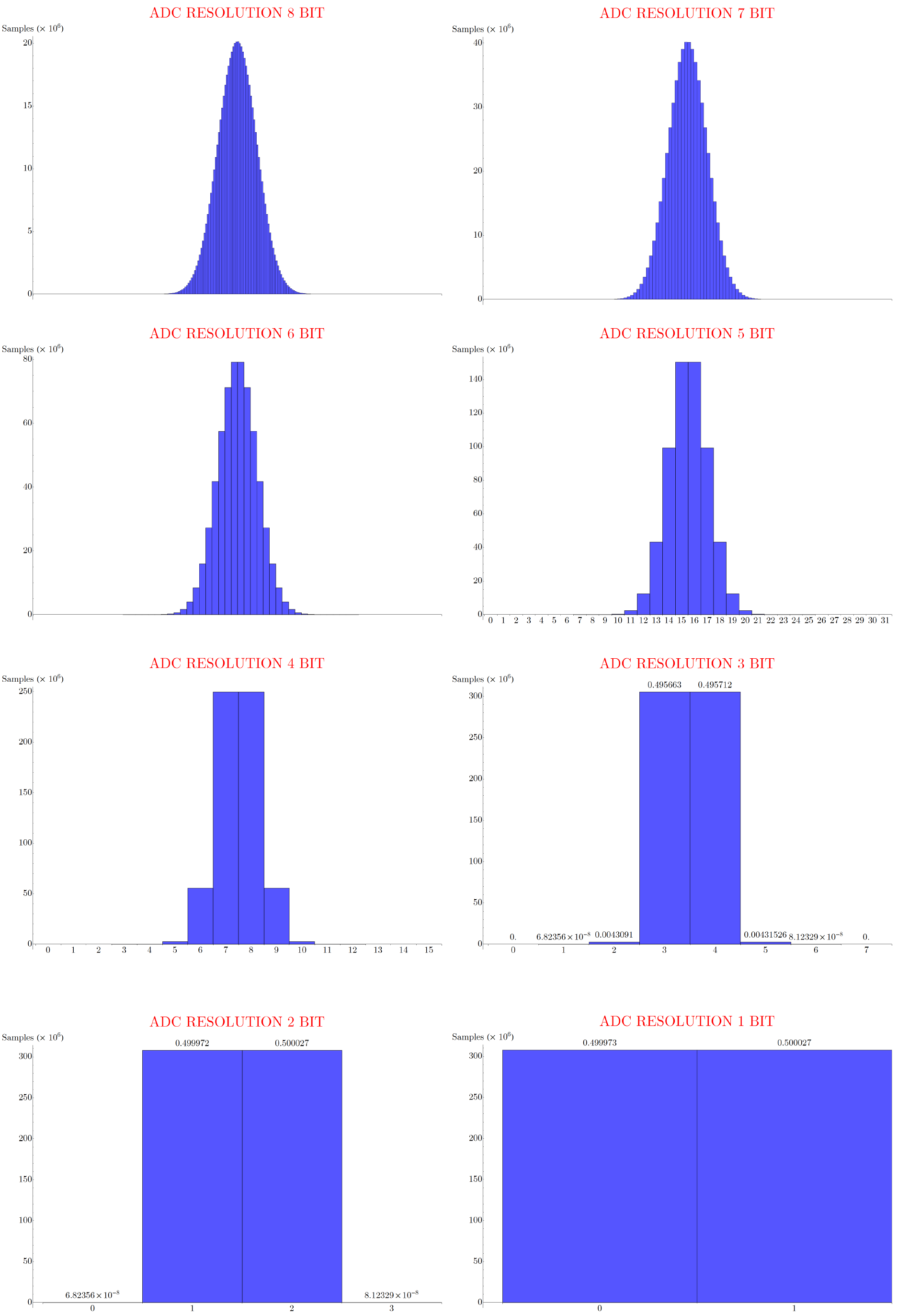}
\end{center}
\caption{In the picture, the various probability distribution obtained by measuring the state with different equivalent ADC resolutions are reported. One notices that the lower the resolution, the more unsharpened is the distribution.}
\label{figSI:risoluzioni}
\end{figure}

\section{Rates}
\begin{figure}[h!]
\begin{center}
\includegraphics[width=\linewidth]{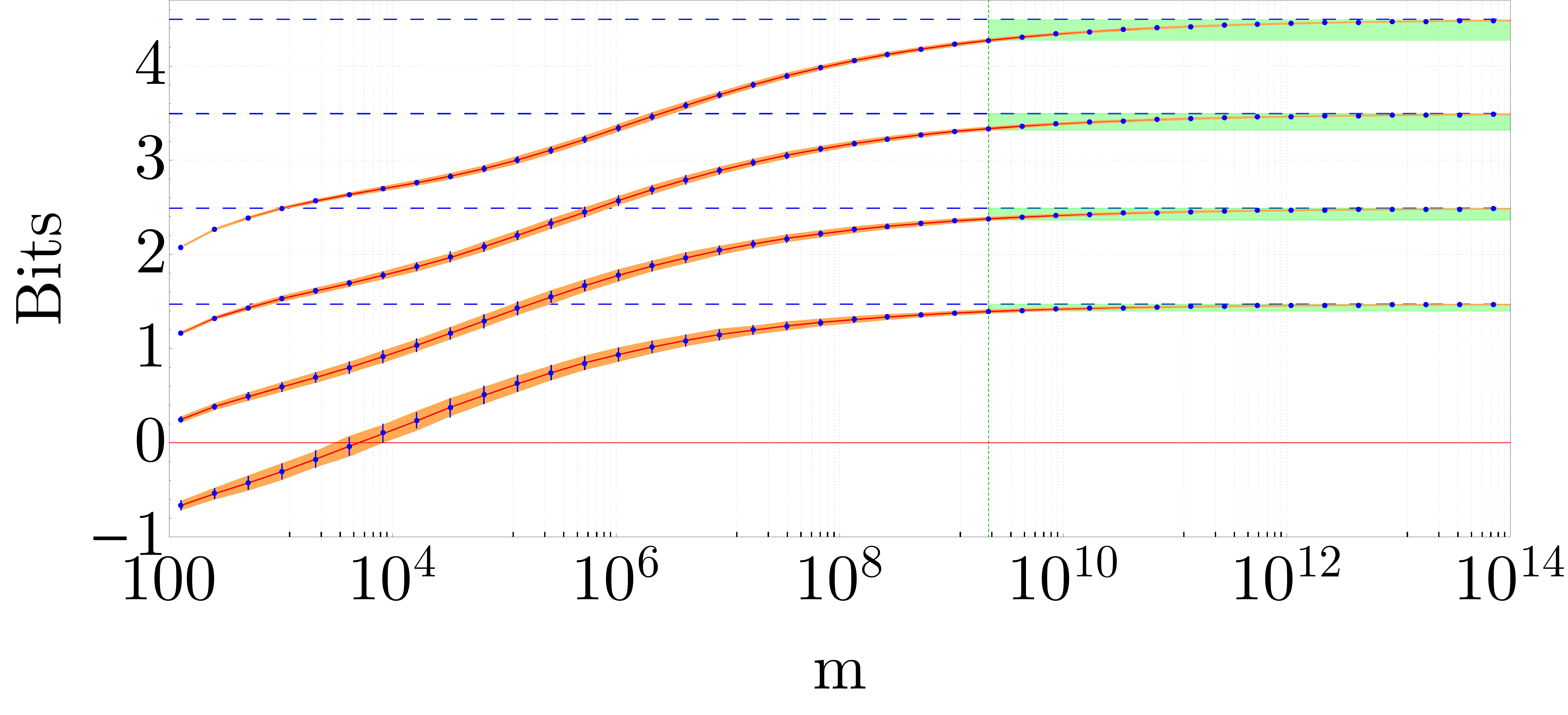}
\end{center}
\caption{In Figure the theoretical and experimental rates, red lines and blue points respectively, are reported for different precisions $\delta_j$ with $j\in\{1,2,3,4\}$ as function of the total number of measurements $m$.  For every $m$, theoretical lines were obtained averaging 100 rate values calculated on simulated sample gaussian distributions with the same variance of the used dataset. The orange shaded areas correspond to expected the 3$\sigma$ errors. The blue points are the averages of the rates evaluated on 200 random samples of size $\lceil \sqrt{m} \rceil$. As one see there is a remarkable agreement between experimental points and expected results. In particular one has that for $m\rightarrow \infty$, the rates tend to an asymptotic value equal to the min conditional entropy. The green dashed line marks the regions where one has less than the 5\% of distance from the asymptotic value.}
\label{figSI:entropie}
\end{figure}

We evaluated the secure generation rate, i.e. the neat number of true random bits per measurement, according
\begin{equation}
r_{\rm sec}=\frac{1}{m}(m-n_Q)[-c(\delta_j)-\widetilde H_{max}(Q_{\delta_j})]-t(m)
\end{equation}
varying the precision $\delta_j$ with $j\in\{0,1,2,3\}$ with the results are represented in Fig. \ref{figSI:entropie}. The rates tend to the asymptotic value of $\widetilde r\xrightarrow{\quad} r(P_{\delta_j})=-c(\delta_j)-\widetilde H_{1/2}(Q_{\delta_j})$ for $m\rightarrow \infty$, cfr. \cite{vallone2014quantum}. The red lines and the orange areas represent the expected average rate and the $3\sigma$ uncertainty respectively, obtained by simulating the check measurement with a gaussian probability distribution having the same measured variance of the sample set. Each blue point with $3\sigma$ error bar, corresponds instead to the averages of $\widetilde r$, being every average evaluated on 200 random data set of size $\sqrt{m}$ with $m\in \{2^7,\dots,2^{47}\}$.  As one can see, there is a remarkable agreement between the expected and the experimental values.

For what concerns a real implementation of the protocol, an advantage in using the highest precision of the ADC,  lies in the fact that one would need much more measurements to reach a given rate value but with a lower precision. The green shaded area in the plot marks the regions where the distance to the asymptotic limit is less than the $5\%$ and which starts at $m=2.1\cdot10^9$: if a QRNG were provided with the equivalent of a squashed quantum state with this same $\sigma^2_Q$, with a sampling rate of 2 GS/s and a full resolution of 8 bits, the generator could provide a quantum secure rate of roughly 8.71 Gbit/s.

\section{Protocol's Noise resiliency}

\begin{figure}[h!]
\begin{center}
\includegraphics[width=0.95\linewidth]{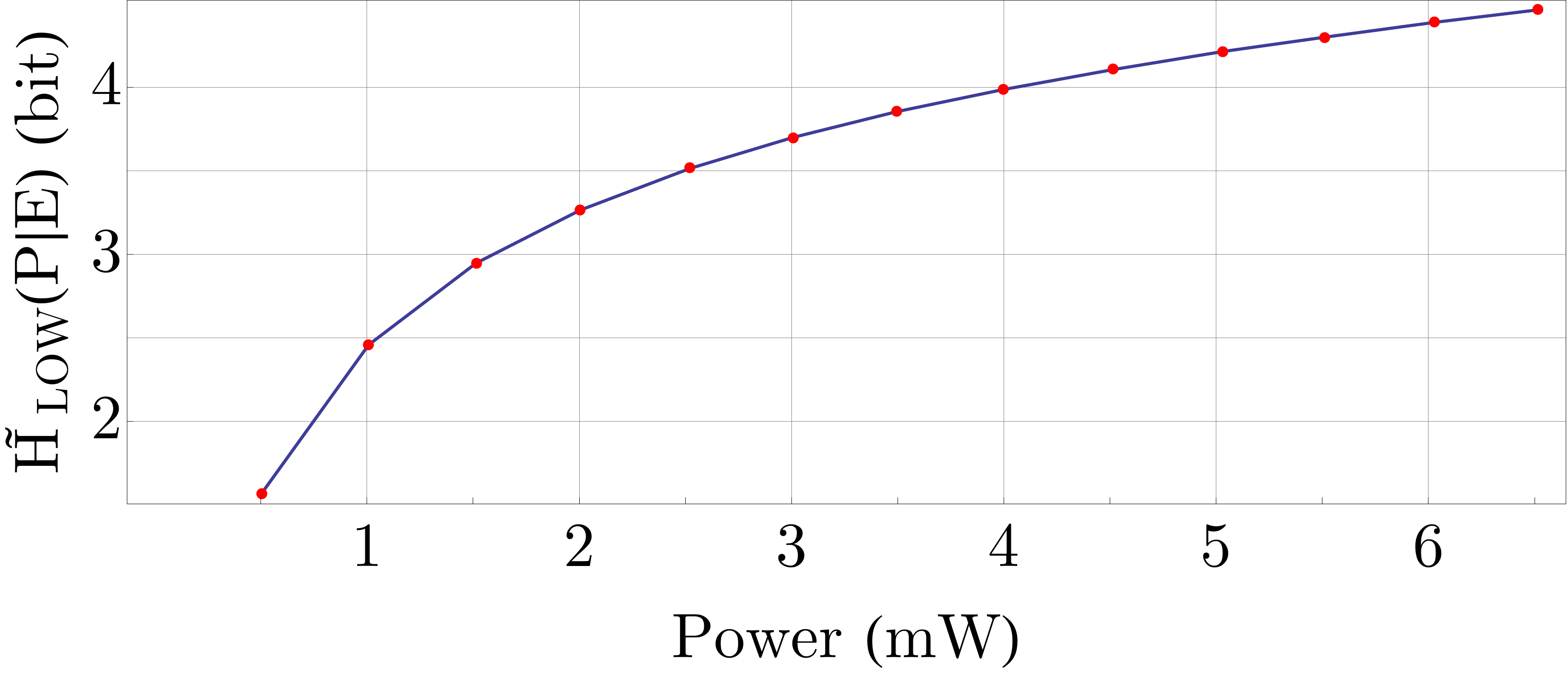}
\end{center}
\caption{We show the value $\widetilde H_{\min}(P_{\delta_0}|E)$ as a function of the power of local oscillator. When the power and then the signal-to-noise ratio (SNR) is higher, the generator yields more secure random bits. On the other hand, at low power, the quantum signal is more affected by the classical electronic noise resulting in a wider thermal-like state. }
\label{snr}
\end{figure}

Our protocol  is dynamically resilient to varying noise conditions because the fraction of extracted true random bits is proportional to the signal to noise (SNR) ratio. In this regard, in Fig. \ref{snr}, the bounds on the entropies are evaluated as function of the LO power. When the power is low, the quantum signal to classical noise ratio is low and correspondingly the thermal signal features a wider variance. Hence, the protocol automatically extracts a lower amount of true random bits.

\section{Results of the statistical tests}

\begin{figure}[ht]
\begin{center}
\includegraphics[width=0.95\linewidth]{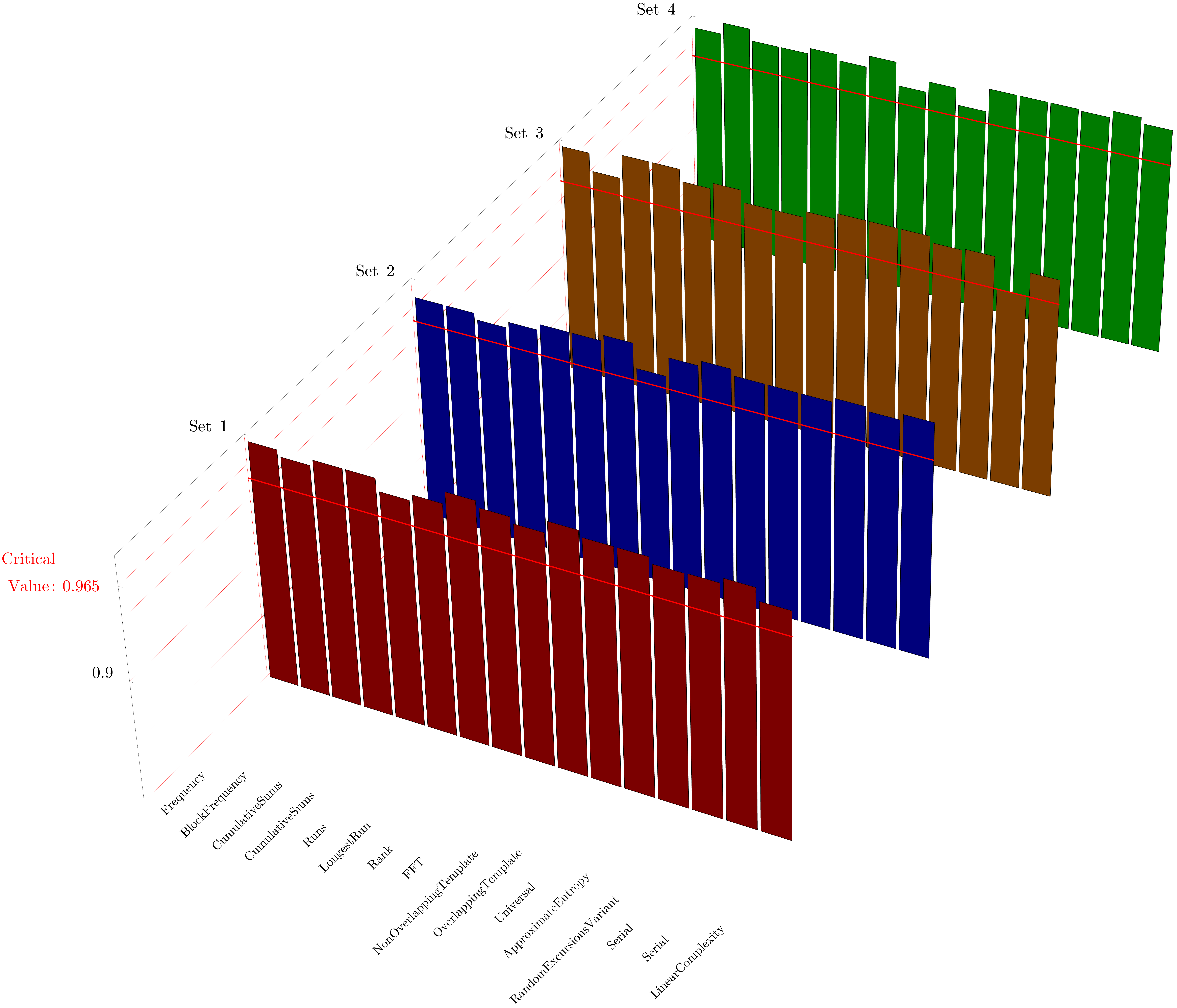}
\end{center}
\label{test_res}
\caption{The stacked plots represents the passing ratio for the NIST test suite SP-800-22. Histograms correspond to four different strings $200^8$ bit long. Each of the 16 tests was applied on 200 substrings $10^6$ bits long and a bin represents the fraction of strings which passed a given test. The red line corresponds to the critical passing ratio.}
\end{figure}

For the post-processing of the numbers, we implemented the fast computable two-universal hash function introduced in \cite{Frauchiger2013}.  The final net total amount of secure random bits from the data set considered in the Main Text, amounted to $8.4 \cdot 10^8$. These were obtained by taking the modulo sum 2 of the product between substrings $n=10000$ bits long and a $n\times l$ random matrix with $l = 2725$ (corresponding to the ratio between $\hat{H}_{\text{LOW}}(P_{\delta p}|E)$ and the binary encoding of a single measurements, i.e. 5 bits). In this proof of principle, the hash matrix was generated for every substring using the pseudo-random number generator of the processing software: naturally a real implementation would require a seed of true random numbers to be stored inside the generator. We tested the numbers with standard NIST statistical tests in order to assess the \emph{statistical} quality of the numbers: it is worth stressing that once the extractor is properly calibrated and the hash matrix is truly random, the post-processed numbers are expected to pass the tests. This was indeed the case as one can see from Fig. \ref{test_res}.

\begin{thebibliography}{10}
\providecommand{\url}[1]{\texttt{#1}}
\providecommand{\urlprefix}{URL }
\providecommand{\eprint}[2][]{\url{#2}}

\bibitem{Schmidt1970}
H.~Schmidt, J. Appl. Phys. \textbf{41}, 462 (1970).

\bibitem{Rarity1994}
J.~G. Rarity, P.~Owens, and P.~Tapster, J. Mod. Opt. \textbf{41}, 2435 (1994).

\bibitem{Jennewein2000}
T.~Jennewein, U.~Achleitner, G.~Weihs, H.~Weinfurter, and A.~Zeilinger, Rev.
  Sci. Instrum. \textbf{71}, 1675 (2000).

\bibitem{stefanov2000optical}
A.~Stefanov, N.~Gisin, O.~Guinnard, L.~Guinnard, and H.~Zbinden, J. Mod. Opt.
  \textbf{47}, 595 (2000).

\bibitem{Furst2010}
M.~F\"{u}rst, H.~Weier, S.~Nauerth, D.~G. Marangon, C.~Kurtsiefer, and
  H.~Weinfurter, Opt. Express \textbf{18}, 13029 (2010).

\bibitem{stipvcevic2007quantum}
M.~Stipcevic and B.~M. Rogina, Rev. Sci. Instrum. \textbf{78}, 45104 (2007).

\bibitem{Wayne2010}
M.~a. Wayne and P.~G. Kwiat, Opt. Express \textbf{18}, 9351 (2010).

\bibitem{Trifonov2007}
A.~Trifonov and H.~Vig, US Pat. 7,284,024  (2007).

\bibitem{Gabriel2010}
C.~Gabriel, C.~Wittmann, D.~Sych, R.~Dong, W.~Mauerer, U.~L. Andersen,
  C.~Marquardt, and G.~Leuchs, Nat. Photonics \textbf{4}, 711 (2010).

\bibitem{Shen2010}
Y.~Shen, L.~Tian, and H.~Zou, Phys. Rev. A \textbf{81}, 063814 (2010).

\bibitem{Symul2011}
T.~Symul, S.~M. Assad, and P.~K. Lam, Appl. Phys. Lett. 2--5 (2011).

\bibitem{Note1}
In fact, universal tests of randomness do not exist.

\bibitem{Robert}
R.~Konig, R.~Renner, and C.~Schaffner, IEEE Trans. Inf. Th. \textbf{55}, 1
  (2009).

\bibitem{Pironio2010b}
S.~Pironio, A.~Ac\'{\i}n, S.~Massar, a.~B. de~la Giroday, D.~N. Matsukevich,
  P.~Maunz, S.~Olmschenk, D.~Hayes, L.~Luo, T.~a. Manning, and C.~Monroe,
  Nature \textbf{464}, 1021 (2010).

\bibitem{Colbeck2012a}
R.~Colbeck and R.~Renner, Nat. Phys. \textbf{8}, 450 (2012).

\bibitem{Gallego2013}
R.~Gallego, L.~Masanes, G.~{De La Torre}, C.~Dhara, L.~Aolita, and
  A.~Ac\'{\i}n, Nat. Commun. \textbf{4}, 2654 (2013).

\bibitem{vallone2014quantum}
G.~Vallone, D.~G. Marangon, M.~Tomasin, and P.~Villoresi, Phys. Rev. A
  \textbf{90}, 52327 (2014).

\bibitem{Frauchiger2013}
D.~Frauchiger, R.~Renner, and M.~Troyer, arXiv:1311.4547 .

\bibitem{Furrer2014}
F.~Furrer, M.~Berta, M.~Tomamichel, V.~B. Scholz, and M.~Christandl, J. Math.
  Phys. \textbf{55}, 122205 (2014).

\bibitem{landau1961prolate}
H.~J. Landau and H.~O. Pollak, Bell Syst. Tech. J. \textbf{40}, 43 (1961).

\bibitem{tomamichel2011leftover}
M.~Tomamichel, C.~Schaffner, A.~Smith, and R.~Renner, Inf. Theory, IEEE Trans.
  \textbf{57}, 5524 (2011).

\bibitem{de2012trevisan}
A.~De, C.~Portmann, T.~Vidick, and R.~Renner, SIAM Journal on Computing
  \textbf{41}, 915 (2012).

\bibitem{renner2011quantum}
R.~Renner, in \emph{Information Theoretic Security}, 52--57, Springer (2011).

\bibitem{Note2}
In order to limit the probability of having outcomes larger than the full-scale
  range, see SI.

\bibitem{Fiorentino2007a}
M.~Fiorentino, C.~Santori, S.~Spillane, R.~Beausoleil, and W.~Munro, Phys. Rev.
  A \textbf{75}, 032334 (2007).

\bibitem{lunghi2015self}
T.~Lunghi, J.~B. Brask, C.~C.~W. Lim, Q.~Lavigne, J.~Bowles, A.~Martin,
  H.~Zbinden, and N.~Brunner, Phys. Rev. Lett. \textbf{114}, 150501 (2015).

\bibitem{mitchell2015strong}
M.~W. Mitchell, C.~Abellan, and W.~Amaya, Phys. Rev. A \textbf{91}, 012314
  (2015).

\bibitem{Eberle2013a}
T.~Eberle, V.~H{\"a}ndchen, J.~Duhme, T.~Franz, F.~Furrer, R.~Schnabel, and
  R.~F. Werner, New Journal of Physics \textbf{15}, 053049 (2013).

\bibitem{bachor2004guide}
H.-A. Bachor and T.~Ralph, WileyVCH 434 (2004).

\bibitem{gray1998photodetector}
M.~B. Gray, D.~A. Shaddock, C.~C. Harb, and H.-A. Bachor, Rev. Sci. Instrum.
  \textbf{69}, 3755 (1998).

\end{thebibliography}
\end{document}